\documentstyle[aas2pp4,psfig]{article}

\def\noi{\noindent}
\def\wisk#1{\ifmmode{#1}\else{$#1$}\fi}
\def\arcpt{\wisk{''\mkern-7.0mu .\mkern1.4mu}}

\def\ha{H\,$\alpha$}
\def\hb{H\,$\beta$}
\def\hg{H\,$\gamma$}

\def\oiii{[O~{\sc iii}]}
\def\oii{[O~{\sc ii}]}

\def\ciii{C~{\sc iii}]}
\def\civ{C~{\sc iv}}

\def\lya{Ly\,$\alpha$}
\def\feii{Fe~{\sc ii}}
\def\nev{[Ne~{\sc v}]}

\def\mgii{Mg~{\sc ii}}
\def\soiv{Si~{\sc iv}}
\def\nv{N~{\sc v}}
\def\heii{He~{\sc ii}}
\def\aliii{Al~{\sc iii}}
\def\nii{[N~{\sc ii}]}
\def\sii{[S~{\sc ii}]}

\def\ew{$W_{\lambda}$}
\def\etal{ et al.}
\def\aopt{$\alpha_{\rm opt}$}

\def\gl{$\lambda$}

\def\kms{km\,s$^{-1}$}

\newcommand{\per}{$^{-1}$}

\newcommand{\hour}{$^{\rm h}$}
\newcommand{\degree}{$^{\circ}$}

\slugcomment{To appear in Ap. J. Supplements}

\lefthead{BAKER ET AL..} 
\righthead{OPTICAL SPECTROSCOPY OF A COMPLETE QUASAR SAMPLE } 

\begin{document}

\title{\normalsize THE MOLONGLO REFERENCE CATALOG 1-JY RADIO SOURCE SURVEY \\
IV: OPTICAL SPECTROSCOPY OF A COMPLETE QUASAR SAMPLE }

\author{\sc Joanne C. Baker}
\affil{\rm School of Physics, University of Sydney, NSW
  2006, Australia and\\ MRAO, Cavendish Laboratory, Madingley Road,
Cambridge CB3 0HE, UK\\ }
\authoremail{jcb@mrao.cam.ac.uk}

\author{\sc Richard W. Hunstead}
\affil{\rm School of Physics, University of Sydney, NSW 2006,
Australia}

\author {\sc Vijay K. Kapahi and C.R. Subrahmanya}
\affil{\rm National Centre for Radio Astrophysics, TIFR, 
        Pune University Campus, Ganeshkhind, Pune -- 411007, India}


\begin{abstract}

Optical spectroscopic data are presented here for quasars from the 
Molonglo Quasar Sample (MQS), which forms part of a complete survey of 1-Jy
radio sources from the Molonglo Reference Catalogue. The combination
of low-frequency selection and complete identifications means that the
MQS is relatively free from the orientation biases which affect most
other quasar samples. To date, the sample includes 105 quasars and 6
BL Lac objects, 106 of which have now been confirmed
spectroscopically. This paper presents a homogenous set of 
low-resolution optical spectra for 79 MQS quasars, the majority of
which have been obtained at the Anglo-Australian Telescope. Full
observational details are given and redshifts, continuum and 
emission-line data tabulated for all confirmed quasars.

\end{abstract}



\begin{center}
\section{INTRODUCTION}
\label{sec:intro}
\end{center}

It has been widely claimed that orientation plays a crucial role in
the classification of active galactic nuclei (AGN), acting to increase the
observed diversity. This idea has been formulated into the 
`unified schemes' for AGN (reviewed by Antonucci 1993) which 
attempt to reduce the diversity by finding evidence that some 
classes of AGN are identical except for viewing direction. One of the
most successful applications has been the description of core- and
lobe-dominated quasars as being identical except for radio-jet orientation.
In this picture, core-dominated quasars are simply foreshortened 
lobe-dominated quasars viewed with Doppler-boosted cores 
(Orr \& Browne 1982; Kapahi \& Saikia 1982).

Implicit in the unified schemes for AGN is the presence of anisotropic
emission in many wavebands. As well as predicting orientation
dependencies, anisotropic emission also implies that selection biases
must affect all samples of AGN to some degree; brighter and therefore
preferentially-oriented AGN will tend to be favoured. To overcome or
compensate for such selection effects, samples should be arguably well
defined (eg.\,Hewett \& Foltz 1994) and if possible selected by some
{\it isotropic\/} property, such as extended radio emission.  However,
additional biases can be introduced by imposing inappropriate flux
limits at other wavelengths, most commonly the optical.  For example,
the paucity of lobe-dominated quasars in the B2 sample compared with
the 3CR (de Ruiter et al. 1986) has been attributed to missing faint
optical counterparts below the Palomar Sky Survey plate limit (Kapahi
\& Shastri 1987). This follows if the optical continuum in quasars is
aspect dependent.

Many observations have pointed to the optical continuum emission in
quasars being highly anisotropic (Browne \& Wright 1985; Jackson \&
Browne 1989).  However, none of these early studies had been able to
disentangle the aspect dependence from the effects of sample
selection. Indeed, with the exception of the 3CR (Laing, Riley
\& Longair 1983), all previous studies have used data from samples with 
unknown biases and selection effects. To rectify this, and also to
quantify the effects of imposing optical constraints on other samples,
we have defined a new sample of southern, radio-loud quasars, the
Molonglo Quasar Sample (MQS), defined below. The MQS has been defined
using minimal selection criteria; it includes all radio-loud quasars
in a region of sky down to a radio flux density limit of 0.95~Jy at 408~MHz. 
Sources were drawn initially from the Molonglo Reference Catalogue
(MRC; Large et al. 1981). The quasar identifications have proceeded 
largely in parallel with identifications of all the other sources 
in the same sky strip, which has ensured that no quasars have 
remained unidentified because they are optically faint, for example, 
or because they have large radio-optical positional offsets as a 
result of asymmetric radio structure.

In this paper, optical spectroscopic data are presented for the MQS,
including tabulated redshifts, continuum and emission-line data.
Optical spectra are shown here for 79 MQS quasars; spectra for the 
remaining quasars will be published elsewhere.  
This paper is the fourth in a series giving basic radio and optical
data for quasars and radio galaxies from the MRC/1-Jy sample.
The radio galaxy identifications are listed in Paper I (McCarthy et al.
1996) and radio images are presented in Paper II (Kapahi et al. 1998a). 
Radio data for the quasars are presented in Paper III (Kapahi et al. 
1998b), together with a detailed description of the 
identification of the quasar sample. A full set of optical finding 
charts is included in Paper III. Other papers based on an earlier
listing of the MQS have been published, including investigations 
into the aspect dependence of the optical continuum in the MQS 
by Baker \& Hunstead (1995) and Baker (1997), and the 
X-ray properties by Baker, Hunstead \& Brinkmann (1995). Many follow-up
projects are underway, including infrared spectroscopy
and intermediate-resolution optical spectroscopy of
high-redshift MQS quasars (Baker \& Hunstead 1996; Baker 1998).


\begin{center}
\section{THE MOLONGLO QUASAR SAMPLE}
\end{center}


As described in a companion paper (Paper III), the selection criteria
for the MQS were designed to minimise the orientation-dependent biases
present in most other radio-selected samples. To achieve this, initial
selection was made at low frequency where the radio emission is
dominated by the steep-spectrum extended components, and complete
optical identifications were then sought.

The complete flux-limited radio sample from which the quasar sample
was selected consists of 557 sources with flux densities exceeding
0.95 Jy at 408 MHz in the MRC in a $10^{\circ}$ declination strip
($-20^{\circ}>\delta>-30^{\circ}$), excluding those with low Galactic
latitude $|b|<20^{\circ}$ and those in the RA range 14\hour 03--20\hour 20
(due to constraints on observing time). 
Within this strip, sources were imaged first at 843~MHz
with the Molonglo Observatory Synthesis Telescope (MOST) (Subrahmanya
\& Hunstead 1986) and then at higher-resolution (1\arcsec) with the
VLA, mostly at 5~GHz (see Paper III). The VLA snapshot images have
allowed the separation of compact and extended radio components from
which the core-to-lobe flux density ratios, $R$, have been estimated.
The values of $R$ at an emitted frequency of 10 GHz, $R_{10}$, are
listed in Paper III for all sources except compact, steep-spectrum
(CSS) sources which we define as having radio linear sizes
$l<20$ kpc ($H_0=50$ \kms\,Mpc$^{-1}$ and $q_0=0.5$ assumed throughout)
and spectral indices  steeper than $\alpha = 0.5$ 
($S \propto \nu^{-\alpha}$) between 408 MHz and 5 GHz; see, for 
example, Fanti \& Fanti (1994).
In common with other studies, $R$ is assumed to be an indicator of
jet-axis orientation (e.g., Orr \& Browne 1982). The distribution of 
$R$ values for the MQS (see Baker 1997) is consistent with the sample 
being randomly oriented apart from an excess of 
core-dominated quasars whose Doppler-enhanced 
emission pushes them above the radio flux limit.
We note that no such bias arises in the optical 
because complete identifications have been made.

Optical counterparts near the radio core and/or radio centroid
positions were identified first by eye (as described in Paper III) by their
`stellar' appearance on the UK Schmidt IIIaJ plates, down to the
limiting magnitude of $b_{\rm J}\approx22.5$ 
(where $b_{\rm J} = B - 0.23(B-V)$, Bahcall \& Soneira 1980).  
Deep CCD imaging (to
$r\sim 24$) of the fields of most of the 557 sources at Las Campanas,
as described in Paper I, also detected a number of
quasar candidates, including some close to the plate limit. 
Spectroscopy was then sought to confirm the identifications.
The MQS comprises a total of 111 quasar candidates, of which
106 have been confirmed spectroscopically to date
(including 6 BL-Lacs, only one with a measured redshift). The
acquisition and reduction of the optical spectra is described in
Section \ref{sec:aatspec}.

\begin{center}
\section{OPTICAL SPECTROSCOPY}
\label{sec:aatspec}
\end{center}

Low-resolution optical spectra have been sought for all MQS quasar
candidates with mostly the 3.9-m Anglo-Australian Telescope (AAT) and also 
the 4-m telescope at the Cerro Tololo Inter-American Observatory (CTIO).  
To date, AAT spectra have been obtained for 77 quasar candidates (plus 2 BL Lac
objects). Another two quasar spectra were obtained with the ESO 3.6m
telescope (Wall \& Shaver, 1993, private communication).  Low
signal-to-noise spectra, confirming only redshifts and
classifications, have been obtained at CTIO for fourteen more quasars
but were inadequate for accurate emission-line measurements and
have not been included in this compilation. Further spectroscopy is
being sought. Another five candidates await spectroscopic
confirmation, including two lobe-dominated and three CSS targets.
Seven more quasars have spectral data published in the literature and
have not been reobserved.  Data for these remaining objects
will be published separately.  The observations and reduction of the
AAT spectra are described below.

\begin{center}
\subsection{\it Journal of Observations}
\end{center}

The majority of spectra were obtained in three observing runs at the
AAT: 1989 August 2--3, 1991 March 15--16 and 1993 June 22--23.  In
addition, a small number of spectra were obtained through AAT service
observations and by collaborators as part of their back-up programs.
Table 1 summarises the instrumental setup and observing conditions for
the AAT observations.

On the AAT, we used the RGO spectrograph at the f/8 Cassegrain focus.
A dichroic beam splitter directed the red light to FORS (Faint Object
Red Spectrograph) and the blue light to either the IPCS (Image Photon
Counting System), or the Blue Thomson (BT) or Tektronix (Tek) CCD
detectors. The spectral resolution of FORS is about 25\,\AA\ FWHM
(10.1\,\AA\ per CCD pixel) over the range 5200--10\,500\,\AA. The blue
detectors were used with the 250B grating, yielding considerably
higher resolution, about 3\,\AA\ per pixel or 7--8\,\AA\ FWHM,
over a range extending from the atmospheric cutoff to the dichroic
cutoff at 5400\,\AA.

\begin{center}
\subsection{\it Observing Strategy}
\end{center}

Observations were made on dark nights with a slit typically
$2$\arcsec\ wide and $2^{\prime}$ long on the sky, oriented at the
parallactic angle.  Differential atmospheric refraction along the slit
from 3500--9000\,\AA\ ranged from $<0$\arcpt 2 for zenith distances
${\rm ZD}<20$\degree\ to $\sim2$\arcsec\ at ${\rm ZD}\sim50$\degree. A narrow
slit was favoured to optimise signal-to-noise ratios, but in cases of
poor seeing ($>2$\arcsec) the slit width was increased to reduce slit
losses (Table 1). Each quasar was observed in consecutive exposures 
with the object in one of two fixed positions on the slit. This 
strategy was used to optimise sky subtraction beneath the object 
spectrum and reduce the effect of local pixel variations.
To minimise atmospheric extinction, observations
were made as close to the zenith as possible.
Standard stars were observed on each night at appropriate airmasses 
to enable flux calibration and to remove atmospheric absorption features. 
Observations of a Cu-Ar comparison lamp were used for wavelength calibration.

\begin{center}
\subsection{\it Spectral Data Reduction Method}
\label{sec:figarored}
\end{center}

The AAT spectra were reduced using the FIGARO package
at the University of Sydney. First, the raw CCD images were
bias-subtracted and divided by a normalised flat-field; this was  
not necessary for the IPCS data.  Sky flats were used in
preference to dome flats as they showed greater uniformity across the
image.  Cosmic ray hits were removed interactively from regions
of the CCD images close to the quasar spectrum, and then automatically over
the whole image; in both cases we used interpolation from neighbouring
pixels. No cosmic ray rejection was required for the IPCS data. 
To maximise the signal-to-noise ratio, spectra were extracted
from the central few pixels of the seeing profile, consistent with the
image spread due to atmospheric dispersion. Wavelength scales for both
the object and sky spectra were set from the Cu-Ar lamp spectrum,
achieving a typical wavelength accuracy of $<$1\,\AA\ rms.

To remove strong telluric absorption bands, the red spectra were divided
by that of a smooth-spectrum star observed at a similar airmass to the
quasar. The strong absorption bands were generally well-subtracted, except
above 9000\,\AA\ where the residuals are magnified by the flux calibration
which is poorly constrained in this region. Corrections were also applied
for atmospheric extinction in the blue in order to recover accurate continuum
slopes; these corrections were mostly negligible except below about
4500\,\AA\ in a few objects observed at large air masses.

Flux calibration was applied using spectrophotometric standard stars
observed on the same night as the quasars and at similar air masses.
Estimated errors in flux density are typically in the range 20--50\%.
The main limitations were from slit losses in poor and variable
seeing, a consequence of the narrow slit used to attain high
signal-to-noise ratios. The loss of narrow-line flux from extended
emission regions is expected to be negligible for the majority of quasar
targets (e.g. Baum et al.\ 1988 and Tadhunter et al.\ 1993). 

After individually extracting and flux-calibrating the FORS and IPCS/Tek
spectra, they were then joined together at wavelengths around 5400\,\AA. 
The blue spectra were re-binned first to match the resolution of FORS, 
improving the signal-to-noise ratio for emission-line measurements, 
particularly in the early IPCS data. 
 In general, the red and blue spectra matched well in
the region of the join. In cases where the match was poor the IPCS/Tek
data were rescaled vertically to the FORS data. 
All the spectra have
been shifted to a heliocentric frame of reference and the wavelength
scale corrected from air to vacuum.  The final wavelength coverage of
the joined spectra is typically 3400--10\,000\,\AA, with an overall
spectral resolution of $\sim$25\,\AA\ FWHM and average signal-to-noise ratio
exceeding ten.

\begin{center}
\subsection{\it Confirmation of Quasar IDs}
\end{center}

After spectroscopy, the original MQS candidate list yielded 100
confirmed quasars and 6 BL Lacs, leaving five still requiring
confirmation. These five quasar candidates are retained in the
list for completeness.

During the ongoing definition of the MQS, a handful of IDs were
excluded at the telescope on the basis of their spectral properties.
These included two targets mis-identified with Galactic stars
and seven objects classified after spectroscopy as narrow-line radio
galaxies. The main spectral requirement for an ID to be classified as
a quasar was the presence of broad emission lines. Stars were readily
eliminated, and galaxies were recognised by their narrow emission
lines or the 4000\,\AA\ break and other characteristic absorption
features.

For the quasars comprising the MQS, the distribution of IIIaJ
Schmidt-plate magnitudes $b_{\rm J}$ (drawn from the COSMOS Southern
Sky Catalogue) is shown in Figure \ref{bjdist}.  The distribution
peaks at magnitudes significantly brighter than the plate limit,
giving confidence that very few faint quasar IDs have been missed and
the sample is relatively unbiased.  Consistent with this, most very
faint new optical counterparts from the deep CCD imaging program have
turned out spectroscopically to be galaxies.

\begin{center}
\section{THE SPECTRA}
\label{sec:spectra}
\end{center}

Seventy-nine observed (AAT plus two ESO) spectra for the MQS are presented
in Figure \ref{spectra}. Quasars with published optical data and
not shown here have references included in Table 2. Also, spectra of 
BL~Lacs (MRC\,B0118$-$272, MRC\,B1309$-$216 and MRC\,B2240$-$260) 
have been left out because
data with much higher signal-to-noise ratios exist in the literature.
Spectra for two borderline quasars/radio galaxies (MRC\,B0032$-$203,
MRC\,B0201$-$214) are included separately in Figure~\ref{bord} --- neither
shows obvious broad lines although the line ratios indicate higher
excitation than is typical for radio galaxies. Until further
investigation, it was decided that these objects should remain in the
galaxy subsample.


\begin{center}
\subsection{\it Comments on Individual Spectra}
\label{sec:indiv}
\end{center}

\noi{\it MRC\,B0029$-$271\/}: A CSS object with an unusual optical spectrum. 
The peak of the strong line at about 8800\,\AA\ would indicate that this 
line is \nii\,\gl6583 rather than \ha\ although the broad base 
may contain \ha. This interpretation is supported by the presence of 
strong \sii\,\gl\gl6717,6734 and weak or absent \hb. The 
low-excitation spectrum is reminiscent of a starburst galaxy, 
or possibly a LINER, although \oiii\ is still relatively strong. 

\noi{\it MRC\,B0030$-$220\/}: The \hb\ and \hg\ lines appear to be 
asymmetric with a tail to the blue. 

\noi{\it MRC\,B0106$-$233\/}: This quasar has an unusual spectrum showing 
copious \feii\ emission and weak \mgii. \oiii\ is very weak.

\noi{\it MRC\,B0123$-$226\/}: The \oiii\ $\lambda4959$/\gl5007 line ratio
appears to be less than $1/3$, perhaps indicating a problem with sky
subtraction or possible blending with \feii\,\gl4924.  The 3000\,\AA\
bump is prominent.  The redshift was first reported by Hunstead et
al.\ (1978); an early spectrum was published by Wilkes et al.\ (1983).

\noi{\it MRC\,B0135$-$247\/}: This spectrum is very similar to that of 
MRC\,B0123$-$226.
The redshift was originally determined by Jauncey et al.\ (1978) 
and spectra have been published by Wilkes et al.\ (1983) and Wilkes (1986).  
Again, \oiii\,$\lambda4959$ appears relatively weak, possibly
as a result of blending with \feii\,\gl4924.

\noi{\it MRC\,B0136$-$231\/}: \lya\ and \civ\ show absorption features just 
blueward of their peaks. This $z_{\rm abs}\approx z_{\rm em}$
absorption is also seen in the noisier spectrum published by Wilkes
(1986).  \heii\ $\lambda1640$ is prominent.

\noi{\it MRC\,B0222$-$234\/}: A marked upturn is present in the continuum blueward 
of \hg\ due to the 3000\,\AA\ bump. 

\noi{\it MRC\,B0222$-$224\/}: A quasar with $z_{\rm abs}\approx z_{\rm em}$ 
absorption in \civ.

\noi{\it MRC\,B0237$-$233\/}: A well-known Gigahertz Peaked Spectrum (GPS)
quasar noted for an apparent overdensity of intervening metal
absorption lines. Previous studies have been made by Wilkes et al.
(1983), Wilkes (1986), Heisler, Hogan \& White (1989) and Foltz et al.\
(1993).  Correction for atmospheric absorption distorts the AAT spectrum
at wavelengths $>9000$\,\AA, and makes the relative flux measurement of
\mgii\ unreliable.

\noi{\it MRC\,B0246$-$231\/}: This is the highest redshift quasar in the 
MQS ($z=2.914$). The continuum is very faint but \lya\ and \civ\ were 
detected clearly.

\noi{\it MRC\,B0327$-$241\/}: The spectrum is very noisy at $\lambda>9000$\,\AA\
making a measurement of the \oiii\ equivalent width impossible. 

\noi{\it MRC\,B0328$-$272\/}: ESO/3.6m spectrum, observed by Wall and Shaver
(1993, private communication). 

\noi{\it MRC\,B0346$-$279\/}: ESO/3.6m spectrum, observed by Wall and Shaver 
(1993, private communication). 

\noi{\it MRC\,B0454$-$220\/}: A bright quasar showing strong broad lines. 
An intervening absorption system corresponding to \mgii\ at $z_{\rm
abs}=0.477$ (Wright et al.\ 1979) is clearly visible to the blue 
of the \mgii\ emission peak; this absorption system has been well studied 
in the optical and UV (e.g., Bergeron \& Kunth 1984; Kinney et al.\ 1985).

\noi{\it  MRC\,B0925$-$203\/}: The blue wing of \mgii\ may be cut off by
absorption, similar to that seen in MRC\,B1244$-$255.

\noi{\it MRC\,B1011$-$282\/}: This bright low-redshift quasar has 
been imaged extensively both in the radio and optical by Hutchings, 
Crampton \& Campbell (1984), Gower \& Hutchings (1984) and Stockton 
\& MacKenty (1987). It shows a nebulous optical extension to the NW,
which appears to align with the stronger radio lobe.
The extended nebulosity has also been studied spectroscopically by
Boisson et al.\ (1994).

\noi{\it MRC\,B1055$-$242\/}: The 3000\,\AA\ bump is very prominent in this
quasar and \mgii\ is notably weak. 

\noi{\it MRC\,B1106$-$227\/}: The \lya\ and \civ\ emission lines show 
absorption just blueward of the line peaks.

\noi{\it MRC\,B1114$-$220\/}: The \lya\ line appears to be heavily absorbed, and 
\civ\ appears to show absorption close to the line peak. Uncertainties in 
flux calibration and sky subtraction longward of 9000\,\AA\ affect the 
relative strength of \mgii, which may also show absorption.

\noi{\it MRC\,B1121$-$238\/}: The \hb\ emission line has a rounded profile 
with no evidence for a narrow component, while \mgii\ shows both a 
narrow and a very broad component. The possible emission feature 
at about 5100\,\AA\ may be accentuated by the join between the FORS 
and IPCS spectra.

\noi{\it MRC\,B1151$-$298\/}: The region around the spectral join at 5400\,\AA\ 
is noisy and so the scaling between the FORS and IPCS spectra is uncertain.

\noi{\it MRC\,B1208$-$277\/}: A faint quasar with poor signal-to-noise 
ratio above 9000\,\AA. \oiii\ and \hb\ are discernible but 
the other lines are weak. 

\noi{\it MRC\,B1212$-$275\/}: Only a FORS spectrum is available --- the broad 
emission line is identified with \mgii. The measurement of \aopt\
is uncertain due to the limited wavelength coverage and presence of the
3000\,\AA\ bump. 

\noi{\it MRC\,B1244$-$255\/}: \mgii\ shows a prominent narrow component. 

\noi{\it MRC\,B1257$-$230\/}: The prominence of the 3000\,\AA\ bump 
makes \aopt\ uncertain.

\noi{\it MRC\,B1301$-$251\/}: The spectrum is noisy above 9000\,\AA, 
making \hb\ and \oiii\ fluxes uncertain.

\noi{\it MRC\,B1309$-$216\/}: A BL-Lac object studied by Blades et al. (1980);
no spectrum is given here. No emission lines were found by Blades et al. 
but a redshift limit of $z\geq 1.49$ was set by the detection of \civ\
absorption lines.

\noi{\it MRC\,B1355$-$236\/}: This quasar was observed on two occasions 
at the AAT, showing little change apart from a possible fall in the
strength of \hb\ between 1991 March and 1993 June. The most recent
observing date only is listed in Table 2, and the corresponding 
spectrum shown in Figure \ref{spectra}. 

\noi{\it MRC\,B2024$-$217\/}: \hb\ is very weak compared with \ha, and appears
to be very broad with no narrow component.  The continuum is also 
very red. 

\noi{\it MRC\,B2035$-$203\/}: \oiii\gl5007 is affected by atmospheric A-band 
absorption.

\noi{\it MRC\,B2111$-$259\/}: The emission lines are very weak in this spectrum; 
the strongest line is assumed to be \mgii\ at $z=0.602$. An emission 
feature just beyond 8000\,\AA\ is consistent with weak \oiii\,\gl5007 at
the same redshift.

\noi{\it MRC\,B2122$-$238\/}: A number of strong absorption lines are 
present in the spectrum; these can be identified as complexes of 
\feii\,\gl\gl2382,2586,2600 and \mgii\,\gl\gl2796,2803 at $z_{\rm abs}
\sim 1.70$ and 1.75.

\noi{\it MRC\,B2156$-$245\/}: A quasar with a steep optical spectrum and weak 
emission lines. \mgii\ may be heavily absorbed.

\noi{\it MRC\,B2158$-$206\/}: Clearly at high-redshift ($z=2.272$), 
this quasar was noted in the sample of
Dunlop et al. (1989) with an incorrect redshift.

\noi{\it MRC\,B2232$-$272\/}: \mgii\ has a broad, flat-topped profile
and \civ\ has an asymmetric profile, possibly due to absorption.
\heii\ $\lambda1640$ is prominent.

\noi{\it MRC\,B2256$-$217\/}: Absorption is evident
at the central emission wavelength of \civ\ and possibly also
\mgii.

\noi{\it MRC\,B2338$-$233\/}: This quasar was observed on two occasions 
(1989 Sept 26 and 1993 June 22). \hb\ and \oiii\ were better recorded 
in the later spectrum shown here in Figure~\ref{spectra}.
The most recent observing date only is listed in Table 2.

\noi{\it MRC\,B2348$-$252\/}: The continuum shape appears to be dominated by 
the 3000\,\AA\ bump and prominent \feii\ emission-line blends;
\aopt, therefore, is uncertain.

\begin{center} 
\section{SPECTROSCOPIC DATA}
\label{sec:obsdata}
\end{center}

Table 2 lists the observational details and optical properties for the
MQS, including observing dates, redshifts, optical spectral indices
and IIIaJ blue magnitudes, $b_{\rm J}$, measured by COSMOS. 
References are supplied for quasars with published redshifts.

\begin{center}
\subsection{\it Redshifts}
\label{sec:redshifts}
\end{center}

Redshifts, listed in Table 2, were measured from the peaks 
of strong emission lines. 
Figure \ref{zdis} shows the distribution of
measured redshifts for the MQS, which span the range
0.0--3.0 (median $z\approx1$). Redshifts have not been established for 
five BL-Lac objects.
Narrow lines were used in preference to broad lines for redshift 
measurements to avoid possible bias arising from systematic differences
(velocity shifts $\sim1000$ km\,s\per) which have been reported
between narrow and broad lines (Gaskell 1982; Tytler \& Fan 1992).
Such velocity shifts were seen in some MQS spectra, and will be
discussed in a separate paper.

\begin{center}
\subsection{\it Optical Spectral Indices}
\label{sec:aopt}
\end{center}

Optical spectral indices, \aopt\ ($f_{\nu}\propto \nu^{-\alpha}$),
have been measured over the observed wavelength range
3400--10\,000\,\AA.  In practice, the observed quasar continuum rarely
follows a simple power law over this range and so large uncertainties
in fitted spectral index result, of the order $\pm 0.2$. One of the
most obvious spectral features which produces a deviation from the
power-law continuum is the so-called `3000\,\AA\ bump', believed to
arise from blended \feii\ and Balmer continuum emission over the
restframe wavelength range 2000--4000\,\AA\ (Oke, Shields \& Korykansky
1984; Wills, Netzer \& Wills 1985).

The MQS includes quasars with a wide range of optical continuum slope,
$-0.3<$\aopt$<3$ with median \aopt$\approx 1$ (Figure \ref{aopt}).
Furthermore, a tail of red quasars appears in the distribution of
\aopt\ in Figure \ref{aopt}.

In Figure \ref{aoptbj} --- a plot of \aopt\ against blue magnitude
$b_{\rm J}$ --- a trend is evident, despite the large scatter,  for red
quasars (steep \aopt) to be optically faint. For example, all MQS quasars with
\aopt$> 1.5$ are fainter than $b_{\rm J}=18$.  The correlation in 
Figure \ref{aoptbj} has a (Kendall's tau) probability of $P=0.001$ 
of occurring by chance (note the correlation remains significant 
for core-dominated quasars alone, $P=0.02$).
The correlation may be viewed alternatively as a paucity of optically-bright, 
red quasars in the MQS, which is surprising since such objects are 
not excluded directly by our selection
criteria.  The fact that red MQS quasars are also optically faint may
argue in favour of reddening of quasar light in these objects.
Interestingly, all the quasars with \aopt$> 1.5$ are lobe-dominated
and CSS quasars, pointing to an intrinsic (possibly
orientation-dependent) origin.  
The trend for lobe-dominated and CSS quasars to be reddened 
more than core-dominated quasars is confirmed by the average
spectral properties of the MQS (Baker \& Hunstead 1995), 
and is consistent with other indicators such as 
Balmer decrements and narrow-line equivalent widths. This issue is 
addressed in a paper by Baker (1997), to which the reader is 
referred.

\begin{center} 
\subsection{\it Emission-Line Fluxes}
\label{sec:emlines}
\end{center}

Restframe emission-line properties have been measured for prominent
lines (see Table 3) using {\sc starlink} DIPSO.  The local
continuum level was fitted in most cases by eye because of the 
difficulty in obtaining an acceptable fit to a power law or 
polynomial over the full wavelength range covered and to take account of
broad emission features, such as the 3000\,\AA\ bump.
Where possible, a polynomial continuum was fitted first as a guide.  
Errors in line fluxes were estimated 
empirically with repeated measurements using extreme
continuum placements.

Broad \hb\ and the narrow \oiii$\lambda\lambda4959,5007$ doublet were
separated by integrating the \oiii\ emission above the red wing of
\hb\ and subtracting it from the total \hb+\oiii\ flux. The \oiii\
doublet itself was not separated for flux measurements because both
lines were often blended at the resolution of FORS.  As the doublet ratio 
should be constant, combining the lines should not introduce
bias.  Due to the low spectral resolution no attempt has been made to separate
contributions to prominent broad lines from blended species (e.g. \lya\
and \nv\ have not been separated, nor have \ha, \nii\ and \sii).
It is recommended that individual spectra be viewed to judge the
uncertainty of line measurements.

Rest-frame fluxes for strong lines are given in Table 3 relative to
\hb\ for low-redshift spectra and relative to \ciii\ at higher
redshifts where \hb\ was not visible. Neither \hb\ nor \ciii\ commonly
suffers absorption in quasar spectra although at low resolution these
lines may be blended with \feii\ and \aliii, respectively (Wills,
Netzer \& Wills 1985; Steidel and Sargent 1991). Because of the low
spectral resolution, we have not attempted to separate the complex
\feii\ blends from the other broad-line emission in the
4500--5400\,\AA\ region.  Line flux ratios are not affected by errors
in the absolute flux calibration; however, uncertainties will be
introduced by differences in the relative scaling of the red and blue
spectra, wavelength-dependent errors in flux calibration and
extinction corrections.  Errors in relative fluxes are large,
estimated to be $\sim30$\%, possibly greater for weak lines or in
regions of poor signal-to-noise.  Neighbouring lines, such as \hb\ and
\oiii\ will have more accurately determined flux ratios.

Generally, the broad lines show a small dispersion in relative fluxes.
On the other hand, Figure \ref{rfluxo} shows that the ratios of \oii\
and \oiii\ to \hb\ flux span nearly four and two decades,
respectively.  The MQS spectra show a range of \ha/\hb\ Balmer
decrements from 3 to 12, suggesting that reddening is important in the
MQS (see Baker 1997).  Line blends may affect the fluxes of both \ha\ and
\hb, although not significantly for strong lines. Other diagnostic
line ratios in the MQS include broad \civ/\ciii\ which ranges between
0.8--10, and narrow \oiii/\oii\ which spans 1--200.

\begin{center}
\subsection{\it Equivalent Widths}
\label{sec:ews}
\end{center}

Equivalent widths, \ew, were measured in the manner described above.
Uncertainties in equivalent widths are typically 5--15\% (see Table 3), and  
are independent of flux calibration. The errors are slightly
larger for the \mgii\ and \oii\ lines due to uncertainty in setting the
continuum level in the (rest frame) 2000--4000\,\AA\ region. Where possible,  
a polynomial was fitted through line-free regions of the spectrum avoiding 
the 3000\,\AA\ bump, and this level was used to calculate \ew. 
The tabulated uncertainties in \ew\ include any differences between
interpolated (polynomial fit) and local continuum flux.

Figure \ref{ewall} shows the distributions of equivalent widths for the
eight most prominent lines in the quasar spectrum: narrow 
\oiii\,\gl\gl4959,5007 and \oii\,\gl3727, broad \ha\,\gl6563,
\hb\,\gl4861, \mgii\,\gl2798, \ciii\,\gl1909, \civ\,\gl\gl1549,1551 and
\lya\,\gl1216. As expected from Figure \ref{rfluxo}, the narrow 
\oii\ and \oiii\ lines show the greatest spread: 4 and 2 decades 
respectively. The small range of \hb\ equivalent widths is notable, 
a consequence of the tight correlation between optical continuum 
and \hb\ line luminosity observed in virtually all types of AGN (Yee 1980).
The distributions of \ew\ in Figure \ref{ewall} for the broad lines
show no prominent asymmetry, except perhaps for \civ. In general, 
the \ew\ ranges measured for the MQS are similar to those found
in other samples of both radio-loud and -quiet quasars
(e.g. Baldwin, Wampler \& Gaskell 1989; Boroson \& Green 1992;
Jackson \& Browne 1991).

\begin{center}
\subsection{\it Emission-Line Widths}
\label{sec:linewidths}
\end{center}

Line widths (FWHM) have been measured for the prominent broad lines
and are listed in Table 3 in \kms.  
No deconvolution of broad and narrow components was attempted in
general; most line profiles were smooth and dominated by the broad
component. In a few cases, however, with a clearly separated 
unresolved component sitting on top of a broad line, 
the width was measured from the broad component alone
using eyeball deconvolution.  Again, the measurement uncertainties are
large, about $20$\%. The widths may be unreliable for emission lines 
with absorption occurring near the line peak. Amongst MQS quasars
with $z>1.4$, where \civ\ and/or \lya\ is visible, absorption systems
within a few thousand \kms\ of the line peak are seen in 
approximately 50\% of the spectra, comparable with other 
studies of steep-spectrum quasars (e.g. Anderson et al. 1987) 
(the occurrence of associated
absorption in the MQS will be addressed elsewhere, see e.g. 
Baker \& Hunstead 1996).
Because a typical emission line profile is rarely Gaussian, as well as
the possibility of absorption, observed FWHMs 
provide only an approximate measure of line width, particularly
for strong lines with very broad wings.  The line profiles and their
aspect dependence will be studied in more detail in a later paper.

Figure \ref{fwhmdis} shows the distribution of velocity widths
(km\,s\per) for the four strongest broad lines. All four lines show a
similar spread in line widths, from 1000--20\,000 km\,s\per. \mgii\
seems to have an asymmetric distribution, with a tail extending to
narrower lines. Again, these values are comparable with velocity widths
for quasars in other samples (e.g. Baldwin, Wampler \& Gaskell 1989).

\begin{center}
\section{NOTES TO TABLES}
\end{center}

The spectroscopic data are presented in Tables 2 and 3. In Table 2, for
each MRC quasar named (B1950 convention) the
observing date is listed in column 2; the source classification
(Q=quasar; Q?=likely quasar; B=BL Lac) in column 3; the redshift in
column 4; the optical spectral index (3000--10000\AA\ observed; $S_{\nu}
\propto \nu^{-\alpha}$) in column 5; and the COSMOS IIIaJ magnitude
(r=$R$-band CCD magnitude) in column 6. References are supplied for
individual quasars in column 7: [1] Blades \etal\ (1980);
[2] Boisson \etal\ (1994);   
[3] Chu \etal\ (1986);             
[4] Dekker \& D'Odorico (1984); 
[5] Dunlop \etal\ (1989);
[6] Gower \& Hutchings (1984);     
[7] Hunstead (1991 priv. comm.);
[8] Hunstead \etal\ (1978);  
[9] Jauncey \etal\ (1978);      [10] Murdoch \etal\ (1984);  
[11] McCarthy (1994 priv. comm.);
[12] O'Dea \etal\ (1991);     
[13] Rawlings (1993 priv. comm.);
[14] Savage \etal\ (1976); 
[15] Stickel \etal\ (1993a);
[16] Stickel \etal\ (1993b); 
[17] Wall \& Shaver (1993 priv. comm.); 
[18] White \etal\ (1988);     
[19] Wilkes et al. (1983), Wilkes (1986);     
[20] Wright \etal\ (1979);  
[21] Wright \etal\ (1983).
An `N' in column 8 denotes additional notes in Section 4, an `S' 
is given if the spectrum is shown in Figure 4. 

Emission line data are presented in Table 3 for the lines
\ha\,\gl6563 ($+$\nii), 
[O\,{\sc iii}]\,\gl\gl4959,5007, \hb\,\gl4861,
\hg\,\gl4340 (blended with \oiii\,\gl4363), \oii\,\gl3727, \nev\,\gl3426, 
\mgii\,\gl2798, \ciii\,\gl1909, \civ\,\gl\gl1549,1551, 
Si~{\sc iv}/O~{\sc iv}]\gl1400 (blend)
and \lya\,\gl1216 (blended \nv\,\gl1240).  
The observed
wavelength of the line peak is given in column 3 (only \gl5007 is given
for the [O\,{\sc iii}] doublet), or references given for data taken from 
the literature:  [1] Hunstead (priv. comm.);
[2] White et al. (1988);  [3] Wilkes (1986). 
Uncertain values are indicated with a
colon throughout the Table. In column 4 the ratio of the 
integrated line flux is given relative to broad \hb\ or \ciii\ 
(when no \hb); letters show if the line measurement was
affected by bad sky subtraction (s), absorption (a), noise (n), 
being near the edge of the spectrum (e) or a join (j), or the line is weak (w).
Restframe 
equivalent widths (\AA) are listed (with their measurement uncertainties)
in column 5; velocity widths (FWHM) are given in column 7 
($\times 10^{3}$\kms) for mostly unblended
permitted lines (forbidden lines are indicated by `u').

\begin{center}
\section{CONCLUSIONS}
\end{center}

Low-resolution optical spectroscopy has been completed for 106/111
quasars (including 6 BL Lacs) which comprise the MQS; 79 quasar
spectra are published here (plus two borderline radio galaxy spectra).  
The high signal-to-noise ratios achieved, even for faint 
($\sim22$\,mag) targets, have allowed measurements of redshifts, 
spectral slopes and a wide range of emission-line properties.
Redshifts for the MQS range from $z\approx 0.1$ to 3, reaching
significantly higher $z$ than the only other completely-identified,
low-frequency-selected sample, the 3CRR (Laing et al. 1983), which
contains only one $z>2$ quasar.  

The MQS quasars show a wide range of spectral properties 
including a large fraction of red quasars, which are mainly
lobe-dominated and CSS quasars. The emission-line properties 
of the MQS are in broad agreement with other studies, showing 
similar ranges in line widths and equivalent widths for the broad lines. 
The narrow oxygen lines, notably \oii, span a particularly wide range in 
equivalent width (see Baker 1997). Further analysis of the MQS 
optical data and follow-up observations are ongoing and will be 
presented in forthcoming papers. 

\newpage

\begin{center}
\section*{ACKNOWLEDGEMENTS} 
\end{center}

We thank the referee, Chris Impey, for his helpful comments. 
JCB was supported for part of this work by a postgraduate scholarship
from the Special Research Centre for Theoretical Astrophysics,
University of Sydney.  RWH acknowledges funding from the Australian
Research Council. Staff at the Anglo-Australian Telescope are thanked
greatly for their support.



\clearpage

\begin{figure}
\caption[Blue Mags]
{Distribution of COSMOS $b_{\rm J}$ magnitudes for the MQS.  The
median $b_{\rm J}\approx19$ is significantly brighter than the IIIaJ 
Schmidt plate limit, $b_{\rm J}\approx 22.5$. }
\label{bjdist}
\end{figure}

\begin{figure}
\caption[AAT]
{AAT optical spectra for the MQS.  Flux density $f_{\nu}$ (mJy) is
plotted as a function of $\lambda$ (\AA). }
\label{spectra}
\end{figure}

\begin{figure}
\caption[borderline]
{Spectra for two MRC sources which were rejected from the MQS as probable
radio galaxies.}
\label{bord}
\end{figure}
 
\begin{figure}
\caption[Redshift Distribution]
{Distribution of measured redshifts for MQS quasars.}
\label{zdis}
\end{figure}

\begin{figure}
\caption[Optical Spectral Index]
{Distribution of optical spectral index, \aopt, for the MQS.  The
overall shape of the distribution is unlikely to be affected
significantly by uncertainties in individual measurements of \aopt\
($20-30$\%) arising from line blends and the 3000\,\AA\ bump.}
\label{aopt}
\end{figure}

\begin{figure}
\caption[Optical Spectral Index vs Blue Mag]
{Plot of optical spectral index, \aopt, versus COSMOS blue magnitude,
$b_{\rm J}$.  Core-dominated quasars ($R>1$) are denoted by `$\bullet$',
lobe-dominated quasars ($R<1$) by `$\odot$', CSS by `$\ast$' and
others by `\large$\circ$\normalsize'. Typical errors are 20--30\% in
\aopt, and 0.3 mag in $b_{\rm J}$.  }
\label{aoptbj}
\end{figure}

\begin{figure}
\caption[Rel Fluxes of oii and oiii]
{Distribution of fluxes of the narrow \oii$\lambda3727$ line and
\oiii$\lambda\lambda4959,5007$ doublet relative
to broad \hb. Both span a large range.}
\label{rfluxo}
\end{figure}

\begin{figure}
\caption[Equivalent Width Dists]
{Equivalent width distributions for eight prominent emission lines;
\oiii\,\gl\gl4959,5007, \oii\,\gl3727, \ha\,\gl6563($+$\nii), \hb\,\gl4861, 
\mgii\,\gl2798, \ciii\,\gl1909, \civ\,\gl1549 and \lya\,\gl1216($+$\nv).
}
\label{ewall}
\end{figure}

\begin{figure}
\caption[FWHMs for Broad Lines]
{Distribution of velocity FWHMs (in km\,s\per) for four strong broad lines:
\hb\,\gl4861, \mgii\,\gl2798, \ciii\,\gl1909 and \civ\,\gl1549.
}
\label{fwhmdis}
\end{figure}


\newpage

\begin{deluxetable}{lcccl}
\tablecolumns{5}
\tablewidth{0pt}
\tablecaption{AAT Observing Log}
\label{obs}
\tablehead{
\colhead{Obs}    & \colhead{Detectors}   & \colhead{average}   
& \colhead{slit} & \colhead{Comments} \\
\colhead{date} &  & \colhead{seeing} & \colhead{width} & \\ 
 & &\colhead{(\arcsec)} & \colhead{(\arcsec)} & }
\startdata
89-08-02 & {\sc fors+ipcs} &1.0 &1.0 &clear  \nl
89-08-03 &     "           &2.0 &1.5 &cloud \nl
89-09-26 &     "           &3.0 &1.5 &poor  \nl
91-03-15 &     "           &2.0 &1.6 &clear  \nl
91-03-16 &     "           &1.0 &1.3 &clear \nl
92-06-04\tablenotemark{a} & {\sc fors+bt}   &2-4 &1.5 &clear  \nl
92-11-29 & {\sc fors+tek}  &2.0 &2.0 &clear  \nl
92-11-30 &    "            &2.0 &2.0 &clear  \nl
93-06-22 &     "           &1.5 &1.7 &cloud  \nl
93-06-23 &   "             &4.5 &2.0 &haze  \nl
93-11-15\tablenotemark{b} & " &1.5 &2.5 &mostly clear  \nl
94-04-16 &   {\sc fors}    &1.7 &1.8 &clear  \nl 
\enddata
\tablenotetext{a}{ATAC Service}
\tablenotetext{b}{Rawlings et al.}
\end{deluxetable}

\newpage


\begin{deluxetable}{ccllrlll}
\small
\textheight 7in
\tablecolumns{8}
\tablewidth{0pt}
\tablecaption{Spectroscopy of Molonglo Quasars}
\label{table_opt}
\tablehead{
\colhead{MRC Quasar} & \colhead{Date} & \colhead{ID} & \colhead{$z$} & 
\colhead{$\alpha_{\rm opt}$} & \colhead{$b_{\rm J}$} & \colhead{Ref} 
& \colhead{Notes} \\
\colhead{(1)} & \colhead{(2)} & \colhead{(3)} & \colhead{(4)} &
\colhead{(5)} & \colhead{(6)} & \colhead{(7)} & \colhead{(8)} }
\startdata 
0017--207 & 93-Jun-22 & Q & 0.545  & 0.38   & 19.3 &7    & S  \nl
0022--297 & 89-Aug-03 & Q & 0.406  & 0.83   & 18.8 &16   & S  \nl
0029--271 & 89-Aug-03 & Q & 0.333  & 1.47   & 19.6 &     & SN \nl
0030--220 & 89-Aug-03 & Q & 0.806  & 1.54   & 18.9 &     & SN \nl
0040--208 & 93-Nov-15 & Q & 0.657  & 0.27   & 16.4 & 13  & S  \nl
0058--229 & 89-Aug-03 & Q & 0.706  & 2.18   & 21.7 &     & S  \nl
0106--233 & 89-Aug-03 & Q & 0.818  & 0.54   & 20.1 &     & SN \nl
0111--256 &           & Q & 1.050  &\nodata & 21.1 &11   &   \nl
0118--272 & 89-Aug-02 & B &\nodata &\nodata & 17.5 &15   &   \nl
0123--226 & 93-Jun-22 & Q & 0.717  & 0.58   & 19.9 &8,19 & SN \nl
0133--266 & 89-Sep-26 & Q & 1.530  & 0.00   & 19.9 &     & S \nl
0135--247 & 93-Jun-23 & Q & 0.835  & 0.31   & 18.9 &9,19 & SN \nl
0136--231 & 92-Nov-30 & Q & 1.895  & 0.50   & 19.7 &19   & SN \nl
0142--278 & 93-Jun-23 & Q & 1.148  & 0.13   & 17.5 &19,21& S   \nl
0209--237 & 89-Aug-02 & Q & 0.680  & 2.07   & 19.0 &     & S \nl
0222--234 & 89-Aug-03 & Q & 0.230  & 1.19   & 18.7 &     & SN \nl
0222--224 & 89-Aug-03 & Q & 1.617  & 1.13   & 19.1 &     & SN \nl
0237--233 & 92-Nov-30 & Q & 2.223  & 0.64   & 16.4 &19   & SN  \nl
0246--231 & 93-Jun-23 & Q & 2.914  & 1.92   & 21.4 &     & SN \nl
0301--243 &           & B&\nodata  &\nodata & 16.4 &11   &  \nl
0315--282 &           & Q & 1.170  &\nodata & 19.9 & 11  &  \nl
0327--241 & 92-Nov-29 & Q & 0.895  & 1.07   & 19.4 &3    & SN \nl
0328--272 & 93-Nov-14 & Q & 1.803  & 1.38   & 18.1 &17   & SN \nl
0338--214 &           & B & 0.048  &\nodata & 16.0 & 11  & \nl
0338--259 &           & Q?&\nodata &\nodata & 22.6r&     & \nl
0338--294 & 89-Aug-03 & Q & 1.139  & 0.98   & 18.9 &     & S\nl
0346--279 & 93-Nov-14 & Q & 0.989  & 1.29   & 20.5 &17,18& SN   \nl
0407--226 &           & Q & 1.480  &\nodata & 21.8r&11   & \nl
0413--210 &           & Q & 0.807  & 1.60   & 18.4 &19   & \nl
0413--296 & 89-Sep-26 & Q & 1.630  & 0.37   & 18.6 &     & S \nl
0418--288 &           & Q & 0.850  &\nodata & 21.1 &11   & \nl
0421--225 &           & Q & 0.362  &\nodata & 17.5 &8    & \nl
0430--278 &           & Q & 1.630  &\nodata & 21.3 &11   & \nl
0437--244 &           & Q & 0.84   &\nodata & 17.5 &11   &  \nl
0439--299 &           & B &\nodata &\nodata & 20.4 &11   & \nl
0447--230 &           & Q & 2.140  &\nodata & 22.0 &11   &   \nl
0450--221 &           & Q & 0.898  & 1.40   & 17.8 &7,8  &  \nl
0451--282 &           & Q & 2.560  & 1.13   & 17.8 &19,21&   \nl
0454--220 & 92-Nov-29 & Q & 0.533  & 0.08   & 18.6 &19,20& SN \nl
0522--215 &           & Q & 1.80   &\nodata & 22.0 &11   &  \nl
0549--213 &           & Q & 2.245  & 0.00   & 19.1 &7,8  &  \nl
0925--203 & 92-Nov-29 & Q & 0.346  & 0.36   & 16.4 &19   & SN \nl
0941--200 & 93-Jun-22 & Q & 0.715  & 0.62   & 17.9 &     & S \nl
1006--299 & 91-Mar-16 & Q & 1.064  & 0.38   & 18.0 &     & S \nl
1010--271 & 91-Mar-16 & Q & 0.436  & 2.13   & 18.9 &     & S \nl
1011--282 & 92-Jun-04 & Q & 0.255  & 0.22   & 16.3 &2    & SN \nl
1017--248 &           & Q?&\nodata &\nodata &\nodata & 11  & \nl
1019--227 &           & Q & 1.55   &\nodata & 21.1 &11   & \nl
1025--229 & 95-Mar-31 & Q & 0.309  & -0.71  & 16.7 &     & S\nl
1025--264 & 91-Mar-15 & Q & 2.665  & 0.89   & 17.5 &     & S\nl
1043--291 & 91-Mar-15 & Q & 2.128  & 1.52   & 18.6 &     & S\nl
1052--272 &           & Q & 1.000  &\nodata &\nodata &11 &  \nl
1055--242 & 91-Mar-15 & Q & 1.090  & 0.95   & 19.9 &     & SN \nl
1106--227 & 91-Mar-16 & Q & 1.875  & 1.54   & 20.8 &     & SN \nl
1114--220 & 91-Mar-15 & Q & 2.282  & 2.77   & 20.2 &     & SN \nl
1117--248 & 93-Jun-23 & Q & 0.462  & 0.71   & 17.3 &14,19& S \nl
1121--238 & 91-Mar-15 & Q & 0.675  & 0.40   & 18.6 &     & SN \nl
1151--298 & 91-Mar-15 & Q & 1.376  & 0.35   & 18.1 &     & SN \nl
1156--221 & 93-Jun-23 & Q & 0.563  & 0.92   & 18.6 &19,20& S \nl
1202--262 & 93-Jun-23 & Q & 0.786  & 1.32   & 19.8 &20   & S \nl
1208--277 & 93-Jun-23 & Q & 0.828  & 1.92   & 18.8 &     & SN \nl
1212--275 & 94-Apr-16 & Q & 1.656  & 0.72   & 19.6 &     & SN \nl
1217--209 & 95-Apr-01 & Q & 0.814  & 0.17   & 20.2 &11   & S\nl
1222--293 & 93-Jun-22 & Q & 0.816  & 0.66   & 18.5 &     & S\nl
1224--262 & 95-Mar-31 & Q & 0.768  & 1.49   & 19.8 &11   & S\nl
1226--297 & 93-Jun-23 & Q & 0.749  & -0.27  & 17.0 &     & S\nl
1232--249 & 93-Jun-23 & Q & 0.352  & 1.17   & 17.0 &19   & S\nl
1244--255 & 93-Jun-22 & Q & 0.635  & 0.27   & 16.2 &14,19& SN  \nl
1247--290 & 94-Apr-16 & Q & 0.770  & 0.43   & 22.1 &     & S\nl
1256--220 & 94-Apr-16 & Q & 1.303  & 0.91   & 19.6 &4    & S\nl
1256--243 & 91-Mar-15 & Q & 2.263  & 1.41   & 17.6 &     & S\nl
1257--230 & 91-Mar-15 & Q & 1.109  & 0.77   & 17.1 &     & SN \nl
1301--251 & 91-Mar-16 & Q & 0.952  & 1.72   & 21.0 &     & SN \nl
1302--208 &           & Q?&\nodata &\nodata & 21.8 &     & \nl
1303--250 & 91-Mar-15 & Q & 0.738  & 0.44   & 17.7 &     & S\nl
1309--216 &      &B&\llap{$>$}1.489&\nodata & 18.5 &1    & N \nl
1311--270 & 92-Jun-23 & Q & 2.186  & 1.15   & 19.3 &18   & S\nl
1327--214 & 92-Jun-04 & Q & 0.528  & 0.03   & 16.4 &     & S\nl
1348--289 &           & Q?&\nodata &\nodata & 19.3 &     & \nl
1349--265 & 93-Jun-23 & Q & 0.934  & 2.50   & 18.4 &     & S \nl
1351--211 & 93-Jun-22 & Q & 1.262  & 1.68   & 18.2 &10   & S\nl
1355--215 & 91-Mar-16 & Q & 1.604  & 1.52   & 19.9 &     & S\nl
1355--236 & 93-Jun-22 & Q & 0.832  & 1.17   & 17.8 &     & SN \nl
1359--281 & 91-Mar-16 & Q & 0.802  & 0.81   & 18.7 &     & S\nl
2021--208 &           & Q & 1.299  & 1.09   & 18.3 &7,10,11& \nl
2024--217 & 93-Jun-23 & Q & 0.459  & 2.50   & 19.1 &18   & SN \nl
2025--206 &           & Q & 1.400  & 0.74   & 18.7 &7    & \nl
2030--230 &           & Q & 0.132  & 2.00   & 19.1 &7,8  & \nl
2035--203 & 93-Nov-15 & Q & 0.516  & 1.14   & 16.4 &13   & SN \nl
2037--234 &           & Q & 1.15   &\nodata & 22.4 &11   & \nl
2040--236 & 93-Nov-15 & Q & 0.704  & 0.21   & 16.8 &13   & S\nl
2059--214 &           & Q?&\nodata &\nodata & 23.1r&     & \nl
2111--259 & 93-Jun-22 & Q & 0.602  & 1.39   & 18.1 &     & SN \nl
2122--238 & 89-Aug-02 & Q & 1.774  & 0.76   & 17.8 &     & SN \nl
2128--208 &           & Q & 1.610  &\nodata & 20.0 &11   & \nl
2136--251 & 93-Jun-23 & Q & 0.940  & 1.97   & 18.1 &     & S \nl
2149--200 & 93-Jun-23 & Q & 0.424  & 1.50   & 17.8 &     & S\nl
2156--245 & 93-Jun-23 & Q & 0.862  & 2.05   & 20.2 &     & SN \nl
2158--206 & 89-Aug-02 & Q & 2.272  & 0.63   & 20.1 &5    & SN \nl
2210--257 & 93-Jun-22 & Q & 1.831  & 1.10   & 17.9 &19,21& S\nl
2211--251 & 89-Aug-02 & Q & 2.508  & 1.59   & 19.6 &     & S\nl
2213--283 & 93-Jun-23 & Q & 0.946  & 0.69   & 16.5 &     & S\nl
2227--214 &           & Q & 1.410  &\nodata & 19.6 &11   & \nl
2232--272 & 89-Aug-02 & Q & 1.495  & 1.20   & 19.5 &     & SN \nl
2240--260 & 89-Aug-03 & B &\nodata &\nodata & 17.9 &15   & \nl
2255--282 & 93-Nov-15 & Q & 0.927  & 0.69   & 16.6 &12,13& S\nl
2256--217 & 89-Aug-02 & Q & 1.779  & 1.31   & 19.9 &     & SN \nl
2257--270 & 93-Jun-23 & Q & 1.476  & 1.02   & 18.3 &19   & S \nl
2338--233 & 93-Jun-22 & Q & 0.715  & 0.51   & 17.3 &     & SN \nl
2338--290 & 93-Jun-22 & Q & 0.446  & 0.76   & 18.2 &     & S\nl
2348--252 & 89-Aug-03 & Q & 1.386  & 1.38   & 17.3 &     & SN \nl
\enddata
\end{deluxetable}


\label{table_opt}

\newpage


\begin{deluxetable}{ccrccr}
\small
\textheight=7in
\tablecolumns{6}
\tablewidth{0pt}
\tablecaption{Emission line data for Molonglo quasars}
\label{ewtab}
\tablehead{
\colhead{Quasar} & \colhead{Line} & \colhead{$\lambda_{\rm obs}$} &
\colhead{$F_{\rm rel}$} & \colhead{\ew}  &
\colhead{$\Delta v$} \\
\colhead{(1)} & \colhead{(2)} & \colhead{(3)} & \colhead{(4)} &
\colhead{(5)} & \colhead{(6)} } 
\startdata
0017$-$207& \oiii\ &7747  & 0.85 &$63\pm8$   &u        \nl
          & \hb\   &7524  & 1.0  &$73\pm16$  &9.5       \nl
          & \hg\   &6741  & 0.17 &$9.4\pm 0.4$ &\nodata    \nl
          & \oii\  &5766  & 0.05 &$2.0\pm 0.4$ &u          \nl
          & \nev\  &5300  & 0.01 &$0.4\pm 0.1$ &u          \nl
          & \mgii\ &4326  & 1.7  &$47\pm  3$   &10.2        \nl \nl
0022$-$297& \ha\   & 9227 & 4.3  &$605\pm 57$  &3.5        \nl
          & \oiii\ & 7039 & 2.3  &$254\pm 18$  &u          \nl
          & \hb\   & 6827 & 1.0  &$103\pm 10$  &4.2        \nl
          & \hg\   & 6112 & 0.4  &$41\pm  3$   &\nodata    \nl
          & \oii\  & 5240 & 0.05 &$4\pm   2$   &u          \nl
          & \nev\  & 4890\rlap{:}& 0.14 &$9\pm  2$   &u     \nl
          & \mgii\ & 3980\rlap{:}& 1.30 &$85\pm 24$  &6.6   \nl \nl
0029$-$271& \sii\  & 8978&\nodata&$20  \pm3 $  &u         \nl
          & \nii\  & 8777&\nodata&$103 \pm10$  &u         \nl
          & \oiii\ & 6676&\nodata&$18  \pm1.5$ &u          \nl \nl
0030$-$220& \oiii\ & 9053 & 1.5  &$114 \pm6$   &u         \nl
          & \hb\   & 8788 & 1.0\rlap{n}  &$80  \pm10$  &5.1    \nl
          & \oii\  & 6725 & 0.35 &$24  \pm2$   &u          \nl
          & \nev\  & 6191 & 0.06 &$3.0 \pm0.6$ &u          \nl
          & \mgii\ & 5122\rlap{:}&1.82  &$111 \pm23$  &7.0        \nl \nl
0040$-$208& \oiii\ & 8289 &0.6   &$41  \pm4$   &u       \nl 
          & \hb\   & 8060 &1.0   &$74  \pm8$   &5.6    \nl
          & \hg\   & 7165 &0.4   &$25  \pm3$   &\nodata      \nl
          & \oii   & 6172 &0.06  &$2.7 \pm0.4$ &u      \nl
          &\mgii   & 4642 &1.9   &$48  \pm5$   &6.3    \nl \nl
0058$-$229& \oiii\ & 8550 &1.5   &$210 \pm6$   &u          \nl
          & \hb\   & 8301 &1.0   &$142 \pm16$  &13.0       \nl
          & \hg\   & 7401 &0.22  &$31  \pm3$   &\nodata          \nl
          & \oii\  & 6362 &0.13  &$18  \pm2$   &u          \nl
          & \nev\  & 5836 &0.13  &$18  \pm2$   &u          \nl
          & \mgii\ & 4868\rlap{:}&0.71\rlap{w}  &$95\pm33$  &8.2      \nl \nl
0106$-$233&\oiii\  & 9102 &0.25  &$15  \pm2$   &u          \nl
          & \hb\   & 8840 &1.0\rlap{n}   &$59  \pm6$   &13.0     \nl
          & \hg\   & 7900 &0.13  &$7   \pm1$   &\nodata          \nl
          & \oii\  & 6771 &0.07  &$2.8 \pm0.4$ &u          \nl
          & \nev\  & 6228 &0.04  &$1.3 \pm0.3$ &u          \nl
          & \mgii\ & 5090\rlap{:}&1.06\rlap{n}  &$28  \pm2$   &13.5   \nl 
          & \ciii\ & 3432\rlap{:}&2.2\rlap{e}   &$39  \pm5$   &17.0    \nl \nl
0123$-$226& \oiii\ & 8626 &0.77  &$52  \pm4$   &u          \nl
          & \hb\   & 8361 &1.0   &$67  \pm18$  &2.8        \nl
          & \hg\   & 7482 &0.37  &$23  \pm3$   &\nodata    \nl
          & \oii\  & 6412 &0.07  &$2.9 \pm0.5$ &u          \nl
          & \nev\  & 5883 &0.16  &$5.8 \pm0.7$ &u          \nl
          & \mgii\ & 4802 &0.28\rlap{a}  &$17  \pm2$   &12.4      \nl \nl
0133$-$266& \oii\  & 9441\rlap{:}&\nodata &\nodata&\nodata     \nl
          & \nev\  & 8654 &\nodata&\nodata&\nodata       \nl
          & \mgii\ & 7087 &0.55  &$106 \pm12$  &9.6       \nl
          & \ciii\ & 4830 &1.0\rlap{n}   &$59  \pm7 $  &2.7   \nl
          & \civ\  & 3920 &3.5   &$124 \pm15$  &13.5       \nl \nl
0135$-$247&\oiii\  & 9182 &0.40  &$34  \pm7 $  &u          \nl
          & \hb\   & 8928 &1.0   &$79  \pm12$  &4.5        \nl
          & \hg\   & 7989 &0.38  &$24  \pm4$   &\nodata     \nl
          & \oii\  & 6836 &0.005 &$0.20\pm0.03$&u          \nl
          & \nev\  & 6278 &0.016 &$0.5 \pm0.1$ &u          \nl
          & \mgii\ & 5144 &0.87  &$22  \pm4 $  &7.9        \nl 
          & \ciii\ & [3]~ &\nodata&(94)&\nodata      \nl \nl
0136$-$231& \mgii\ & 8112 &0.94  &$20  \pm4$   &2.7    \nl
          & \ciii\ & 5520 &1.0   &$13  \pm2$   &5.0       \nl
          & \civ\  & 4498 &7.2\rlap{a}   &$60  \pm7$   &8.3  \nl
          & \soiv\ & 4088 &3.0   &$21  \pm2$   &\nodata         \nl
          & \lya\  & 3519 &16.2\rlap{a}  &$94  \pm9$   &10.9   \nl \nl
0142$-$278& \oii\  & 8002\rlap{:}&0.004 &$0.2 \pm0.1$ &u          \nl
          & \nev\  & 7366\rlap{:}&0.011 &$0.5 \pm0.1$ &u          \nl
          & \mgii\ & 6031 &0.33  &$11  \pm2$   &4.2        \nl
          & \ciii\ & 4103 &1.0   &$18  \pm1$   &9.1       \nl \nl
0209$-$237& \oiii\ & 8406 &2.65  &$294 \pm23$  &u          \nl
          & \hb\   & 8163 &1.0   &$104 \pm24$  &4.8        \nl
          & \hg\   & 7290 &0.38  &$46  \pm6$   &\nodata      \nl
          & \oii\  & 6269 &0.54  &$72  \pm9$   &u          \nl
          & \nev\  & 5753 &0.18  &$24  \pm3$   &u          \nl
          & \mgii\ & 4765\rlap{:}&1.05\rlap{w}  &$210 \pm78$  &9.6   \nl \nl
0222$-$234& \ha\   & 8078 &5.1   &$560 \pm26$  &4.2       \nl
          &\oiii\  & 6161 &0.91  &$76  \pm4 $  &u          \nl
          & \hb\   & 5977 &1.0   &$83  \pm4 $  &9.3       \nl
          & \hg\   & 5364\rlap{:}&0.62  &$50  \pm4$   &\nodata  \nl
          & \oii\  & 4592 &0.21  &$13  \pm2$   &u       \nl
          & \nev\  & 4223 &0.07  &$3.0 \pm0.5$ &u       \nl \nl
0222$-$224& \oii\  & 9703 &0.35  &$16  \pm3$   &u       \nl
          & \mgii\ & 7277 &0.79\rlap{a}  &$24  \pm5$   &8.8        \nl
          & \ciii\ & 5031 &1.0\rlap{n}   &$22  \pm1$   &5.8       \nl 
          & \civ\  & 4080 &5.22\rlap{a}  &$88  \pm15$  &18.6    \nl \nl
0237$-$233& \mgii\ & 9057\rlap{:}&0.19\rlap{e}  &$7 \pm3$   &2.5       \nl
          & \ciii\ & 6157 &1.0   &$19  \pm2$   &7.5     \nl
          & \civ\  & 4990 &1.56  &$19  \pm2$   &7.5       \nl
          & \soiv\ & 4514 &0.30  &$4.0 \pm0.5$ &\nodata         \nl
          & \lya\  & 3921 &6.1\rlap{a}   &$75  \pm7$   &16.0    \nl \nl
0246$-$231& \ciii\ & 7487 &1.0\rlap{n}   &$12  \pm5$   &2.3     \nl
          & \civ\  & 6050 &2.51  &$31  \pm7$   &3.7      \nl
          & \soiv\ & 5491 &0.95\rlap{n}  &$14  \pm2$   &\nodata  \nl
          & \lya\  & 4771 &7.5   &$110 \pm15$  &7.6       \nl \nl
0327$-$241&\oiii\  & 9487 &\nodata&\nodata&\nodata    \nl
          & \hb\   & 9207 &1.0\rlap{s}   &$40  \pm6$   &2.7     \nl
          & \hg\   & 8229 &0.4   &$15  \pm2$   &\nodata          \nl
          & \oii\ &\nodata&$<0.1$&$<0.8$ &u         \nl
          & \nev\  & 6476 &0.2   &$3   \pm1$   &u          \nl
          & \mgii\ & 5314 &10.7\rlap{w}  &$74  \pm8$   &4.9   \nl  
          & \ciii\ & 3612 &9.7\rlap{e}   &$46  \pm8$   &\nodata     \nl \nl
0328$-$272& \ciii  & 5351 &1.0   &$31  \pm4$   &4.6      \nl 
          & \civ   & 4341 &2.5   &$63  \pm7$   &4.9   \nl \nl
0338$-$294& \hg\   & 9313\rlap{:}&0.33  &$45  \pm2$   &\nodata    \nl
          & \oii\  & 7948 &0.09  &$8.6 \pm0.7$ &u          \nl
          & \nev\  & 7315 &0.07  &$3.0 \pm0.3$ &u          \nl
          & \mgii\ & 5961 &0.97  &$58  \pm4 $  &10.9       \nl
          & \ciii\ & 4121 &1.0   &$45  \pm4 $  &7.9       \nl 
          & \civ\  & 3338 &3.2\rlap{e}   &$200 \pm100$ &7.0  \nl \nl
0346$-$279& \mgii  & 5569 &1.2   &$22  \pm2  $ &2.4    \nl    
          & \ciii  & 3790 &1.0   &$15  \pm2  $ &4.7    \nl \nl
0413$-$210& \mgii\ & [3]~ &\nodata&(64)&\nodata       \nl
          & \ciii\ &\nodata&\nodata&(38)&\nodata          \nl \nl
0413$-$296& \mgii\ & 7325 &0.75  &$33  \pm2$   &8.6        \nl
          & \ciii\ & 5024 &1.0   &$23  \pm3$   &6.8       \nl
          & \civ\  & 4088 &3.33  &$48  \pm4$   &12.8      \nl \nl
0450$-$221& \mgii\ & [1]~ &\nodata&(19)&\nodata         \nl \nl
0451$-$282& \civ\  & [3]~ &\nodata&(95)&\nodata       \nl
          & \lya\  & [3]~ &\nodata&(264)&\nodata         \nl \nl
0454$-$220& \ha\   &10050 &4.5   &$240 \pm26$  &4.8       \nl
          & \oiii\ & 7694 &0.35  &$21  \pm2 $  &u         \nl
          & \hb\   & 7458 &1.0   &$58  \pm6 $  &6.2       \nl
          & \hg\   & 6681 &0.11  &$5.5 \pm0.5$ &\nodata     \nl
          & \oii\  & 5714 &0.004 &$0.13\pm0.07$&u          \nl
          & \nev\  & 5251 &0.027 &$0.7 \pm0.07$&u          \nl
          & \mgii\ & 4302 &2.24\rlap{a}  &$40  \pm4 $  &11.7   \nl \nl
0549$-$213& \ciii\ & [1]~ &\nodata&(24)&\nodata        \nl
          & \civ\  & [1]~ &\nodata&(57)&\nodata         \nl
          & \lya\  & [1]~ &\nodata&(73)&\nodata         \nl \nl
0925$-$203& \ha\   & 8845 &4.5   &$327 \pm28$  &3.2        \nl
          &\oiii\  & 6744 &0.26  &$24  \pm2 $  &u           \nl
          & \hb\   & 6556 &1.0   &$58  \pm5 $  &2.8        \nl
          & \hg\   & 5847 &0.47  &$28  \pm2 $  &\nodata      \nl
          & \oii\  & 5019 &0.02  &$0.6 \pm0.1$ &u          \nl
          & \nev\  & 4609 &0.03  &$0.9 \pm0.1$ &u          \nl
          & \mgii\ & 3777 &1.45\rlap{a}  &$30  \pm3 $  &7.0    \nl \nl
0941$-$200& \oiii\ & 8605 &0.71  &$35  \pm3 $  &u        \nl
          & \hb\   & 8351 &1.0   &$49  \pm6 $  &5.6      \nl
          & \hg\   & 7472 &0.49  &$23  \pm4 $  &\nodata    \nl
          & \oii\  & 6402 &0.11  &$3.6 \pm0.6$ &u          \nl
          & \nev\  & 5882 &0.02  &$0.9 \pm4 $  &u          \nl
          & \mgii\ & 4802 &1.29  &$28  \pm3 $  &11.0       \nl \nl
1006$-$299& \hg\   & 8987 &0.13  &$8   \pm1 $  &\nodata     \nl
          & \oii\  & 7720 &0.02  &$1.0 \pm0.2$ &u          \nl
          & \nev\  & 7094 &0.025 &$0.9 \pm0.2$ &u         \nl
          & \mgii\ & 5777 &1.58  &$41  \pm4 $  &10.7       \nl
          & \ciii\ & 3919 &1.0   &$10  \pm1 $  &7.1       \nl \nl
1010$-$271& \ha\   & 9426 &7.8   &$563 \pm58$  &5.8     \nl
          &\oiii\  & 7187 &2.0   &$143 \pm15$  &u         \nl
          & \hb\   & 6978 &1.0   &$71  \pm7 $  &8.3       \nl
          & \hg\   & 6248 &0.26  &$19  \pm2 $  &\nodata    \nl
          & \oii\  & 5356 &0.30  &$21  \pm2 $  &u          \nl
          & \nev\  & 4915 &0.14  &$10  \pm1 $  &u          \nl
          & \mgii\ & 4019 &0.8   &$56  \pm6 $  &11.6       \nl \nl
1011$-$282& \ha\   & 8230 &3.6   &$488 \pm50$  &3.9       \nl
          & \oiii\ & 6283 &0.73  &$61  \pm6 $  &u          \nl
          & \hb\   & 6100 &1.0   &$82  \pm8 $  &4.8        \nl
          & \hg\   & 5450 &0.38  &$29  \pm3 $  &\nodata    \nl
          & \oii\  & 4683 &0.032 &$1.6 \pm0.1$ &u          \nl
          & \nev\  &\nodata&$<0.02$&$<1$&u        \nl
          & \mgii\ & 3476\rlap{:}&2.56  &$35  \pm3$   &7.5        \nl \nl
1025$-$229& \oiii\ & 6557 &0.5   &$35  \pm3$   &u          \nl
          & \hb\   & 6368 &1.0   &$67  \pm7$   &9.6      \nl
          & \hg\   & 5690 &0.4   &$24  \pm2$   &\nodata      \nl
          & \oii\  & 4878 &0.03  &$0.7 \pm0.2$ &u          \nl
          & \nev\  & 4491 &0.01  &$0.3 \pm0.1$ &u       \nl
          & \mgii\ & 3663 &4.6\rlap{e}   &$66  \pm7 $  &10.2     \nl \nl
1025$-$264& \ciii\ & 7006 &1.0   &$25  \pm3 $  &6.0      \nl
          & \civ\  & 5673 &1.3\rlap{j}   &$24  \pm3 $  &6.2       \nl
          & \soiv\ & 5150 &1.5   &$17  \pm2 $  &\nodata         \nl
          & \lya\  & 4455 &10.9  &$113 \pm12$  &12.3       \nl \nl
1043$-$291& \mgii\ & 8751 &1.17  &$30  \pm3 $  &3.7        \nl
          & \ciii\ & 5947 &1.0   &$23  \pm2 $  &4.7       \nl
          & \civ\  & 4849 &5.2   &$73  \pm8 $  &6.8       \nl
          & \soiv\ & 4389 &1.2   &$19  \pm2 $  &\nodata    \nl
          & \lya\  & 3802 &12.4  &$128 \pm14$  &7.9       \nl \nl
1055$-$242& \hg\   & 9057\rlap{:}&0.09  &$2.0 \pm0.4$ &\nodata      \nl
          & \oii\  & 6556 &0.20  &$3   \pm0.5$ &u         \nl
          & \nev\  & 7164 &0.02  &$0.2 \pm0.05$&u         \nl
          & \mgii\ & 5849 &0.6\rlap{w}   &$6 \pm2 $  &6.1        \nl
          & \ciii\ & 3991 &1.0\rlap{w}   &$10 \pm1$  &7.5        \nl \nl
1106$-$227& \mgii\ & 8049 &0.78  &$36  \pm4 $  &3.1        \nl
          & \ciii\ & 5470 &1.0   &$31  \pm4 $  &4.7       \nl
          & \civ\  & 4456 &1.20\rlap{a}  &$27  \pm3 $  &6.4       \nl
          & \soiv\ & 4012\rlap{:}&0.38\rlap{a}  &$8 \pm1$   &\nodata      \nl
          & \lya\  & 3497 &7.54\rlap{a}  &$96  \pm11$  &7.4       \nl \nl
1114$-$220& \mgii\ & 9206 &0.86\rlap{a}  &$17  \pm2 $  &6.5        \nl
          & \ciii\ & 6264 &1.0\rlap{a}   &$20  \pm2 $  &8.2      \nl
          & \civ\  & 5081 &1.74\rlap{a}  &$44  \pm4 $  &8.3      \nl
          & \lya\  & 4061 &\nodata&\nodata&\nodata  \nl \nl
1117$-$248& \ha\   & 9611 &3.49  &$187 \pm20$  &5.2      \nl
          & \oiii\ & 7344 &1.05  &$46  \pm6 $  &u          \nl
          & \hb\   & 7122 &1.0   &$42  \pm5 $  &5.6       \nl
          & \hg\   & 6370 &0.49  &$19  \pm2 $  &\nodata    \nl
          & \oii\  & 5459 &0.12  &$3.5 \pm0.3$ &u          \nl
          & \nev\  & 5014 &0.014 &$0.4 \pm0.1$ &u          \nl
          & \mgii\ & 4093 &1.0   &$21  \pm4 $  &10.0       \nl \nl
1121$-$238& \oiii\ & 8312 &1.33  &$80  \pm8 $  &u          \nl
          & \hb\   & 8120 &1.0   &$60  \pm6 $  &8.5       \nl
          & \hg\   & 7285 &0.19  &$11  \pm2 $  &\nodata     \nl
          & \oii\  & 6243 &0.17  &$7   \pm1 $  &u          \nl
          & \nev\  & 5739 &0.09  &$1.3 \pm1 $  &u          \nl
          & \mgii\ & 4683 &4.0   &$80  \pm8 $  &18.1       \nl \nl
1151$-$298& \oii\  & 8866 &0.03  &$1.1 \pm0.3$ &u          \nl
          & \nev\  & 8130 &0.01  &$0.25\pm0.05$&u          \nl
          & \mgii\ & 6644 &1.7   &$28  \pm3 $  &6.1        \nl
          & \ciii\ & 4535 &1.0   &$18  \pm2 $  &5.8       \nl 
          & \civ\  & 3861 &5.6   &$45  \pm5 $  &8.7      \nl \nl
1156$-$221& \oiii\ & 7842 &4.0   &$245 \pm30$  &u        \nl
          & \hb\   & 7620 &1.0   &$61  \pm7 $  &4.8        \nl
          & \hg\   & 6794 &0.63  &$37  \pm6 $  &\nodata          \nl
          & \oii\  & 5830 &1.0   &$43  \pm7 $  &u          \nl
          & \nev\  & 5343 &0.06  &$2.3 \pm0.3$ &u          \nl
          & \mgii\ & 4389 &1.78  &$68  \pm13$  &12.2       \nl \nl
1202$-$262& \oiii\ & 8930 &0.42\rlap{s}  &$54  \pm8 $  &u        \nl
          & \hb\   & 8707 &1.0   &$130 \pm18$  &7.5       \nl
          & \hg\   & 7772 &0.33  &$37  \pm6 $  &\nodata      \nl
          & \oii\ &\nodata&$<0.01$&$<0.5$&u       \nl
          & \nev\  & 6115 &0.017 &$1.3 \pm0.2$ &u          \nl
          & \mgii\ & 5010 &1.27  &$76  \pm6  $ &6.9        \nl \nl
1208$-$277& \oiii\ & 9145 &1.8   &$290 \pm50 $ &u          \nl
          & \hb\   & 8774\rlap{:}&1.0\rlap{n}   &$145 \pm50$  &6.5      \nl
          & \hg\   & 7927 &0.75\rlap{n}  &$109 \pm30$  &\nodata    \nl
          & \oii\  & 6815 &0.13  &$18  \pm2 $  &u          \nl
          & \nev\  & 6296\rlap{:}&0.06  &$9   \pm1$   &u          \nl
          & \mgii\ & 5109\rlap{:}&0.28  &$50  \pm5$   &5.5        \nl \nl
1212$-$275& \mgii\ & 7406 &\nodata&$8.4 \pm1.0$ &3.6          \nl \nl
1217$-$209& \mgii\ & 5084 &\nodata &$44  \pm4  $ &9.7     \nl \nl
1222$-$293& \oiii\ & 9103 &1.63  &$411 \pm38 $ &u          \nl
          & \hb\   & 8827 &1.0   &$252 \pm30 $ &4.9        \nl
          & \hg\   & 7916 &0.27  &$55  \pm8  $ &\nodata    \nl
          & \oii\  & 6783 &0.13  &$18  \pm3  $ &u          \nl
          & \nev\  & 6232 &0.11  &$13  \pm2  $ &u          \nl
          & \mgii\ & 5088 &2.14  &$178 \pm24 $ &8.6        \nl \nl
1224$-$262& \oiii\ & 8853 &4.6   &$390 \pm50 $ &u          \nl
          & \hb\   & 8629 &1.0\rlap{s}   &$86  \pm20 $ &4.8   \nl
          & \hg\   & 7724 &1.0   &$86  \pm15 $ &\nodata          \nl
          & \oii\  & 6603 &1.3   &$111 \pm16 $ &u          \nl
          & \nev\  & 6066 &0.16  &$10  \pm1  $ &u          \nl 
          & \mgii\ & 4963 &2.0   &$107 \pm20 $ &4.6        \nl \nl
1226$-$297& \oiii\ & 8764 &0.2   &$14  \pm3  $ &u          \nl
          & \hb\   & 8520 &1.0   &$72  \pm12 $ &8.3       \nl
          & \hg\   & 7630 &0.24  &$44  \pm9  $ &\nodata      \nl
          & \oii\  & 6518 &0.02  &$1.0 \pm0.3$ &u          \nl
          & \nev\  & 5988 &0.03  &$1.4 \pm0.3$ &u          \nl
          & \mgii\ & 4898 &1.6   &$30  \pm4  $ &9.5        \nl \nl   
1232$-$249& \ha\   & 8891 &7.7   &$766 \pm68 $ &4.5       \nl
          & \oiii\ & 6793 &3.6   &$336 \pm29 $ &u          \nl
          & \hb\   & 6592 &1.0   &$95  \pm14 $ &6.5       \nl
          & \hg\   & 5894 &0.34  &$30  \pm4  $ &\nodata          \nl
          & \oii\  & 5052 &0.21  &$17  \pm2  $ &u           \nl
          & \nev\  & 4648 &0.12  &$9   \pm1  $ &u          \nl
          & \mgii\ & 3793 &0.55\rlap{w}  &$46  \pm8  $ &8.2        \nl \nl
1244$-$255& \oiii\ & 8203 &0.73  &$52  \pm4  $ &u          \nl
          & \hb\   & 7959 &1.0   &$64  \pm6  $ &4.6        \nl
          & \hg\   & 7122 &0.41  &$24  \pm2  $ &\nodata     \nl
          & \oii\  & 6095 &0.05  &$1.8 \pm0.4$ &u          \nl
          & \nev\  & 5607 &0.03  &$0.9 \pm0.2$ &u          \nl
          & \mgii\ & 4580 &1.2\rlap{a}   &$24  \pm3  $ &12.3    \nl \nl
1247$-$290& \oiii\ & 8863 &1.23  &$340 \pm60 $ &u          \nl
          & \hb\   & 8609 &1.0\rlap{n}   &$240 \pm60 $ &6.2    \nl
          & \hg\   & 7681 &0.04  &$8   \pm1  $ &\nodata      \nl
          & \oii\  & 6600 &0.006 &$0.7 \pm0.1$ &u          \nl
          & \nev\  & 6061 &0.02  &$3   \pm1  $ &u          \nl \nl
1256$-$220& \mgii\ & 6446 &\nodata  &$26  \pm3 $  &4.7        \nl \nl
1256$-$243& \mgii\ & 9134 &0.27  &$14  \pm2 $  &3.6        \nl
          & \ciii\ & 6211 &1.0   &$30  \pm3 $  &8.2       \nl
          & \civ\  & 5053 &1.21  &$37  \pm3 $  &9.1       \nl
          & \soiv\ & 5065 &0.3   &$9   \pm1 $  &\nodata      \nl
          & \lya\  & 3974 &5.56  &$161 \pm17$  &14.8       \nl \nl
1257$-$230& \hg\   & 9173 &0.61  &$25  \pm3 $  &\nodata       \nl
          & \oii\  & 7874 &0.06  &$2.3 \pm0.5$ &u          \nl
          & \nev\  & 7225 &0.06  &$1.5 \pm0.5$ &u          \nl
          & \mgii\ & 5893 &1.5   &$28  \pm3 $  &8.9        \nl
          & \ciii\ & 4016 &1.0   &$12  \pm1 $  &6.1         \nl \nl
1301$-$251& \oiii\ & 9782 &2.38\rlap{s}  &$218 \pm25$  &u       \nl
          & \hb\   & 9488\rlap{:}&1.0\rlap{s}   &$92 \pm14$  &5.8    \nl
          & \hg\   & 8482 &0.34  &$31  \pm6 $  &\nodata          \nl
          & \oii\  & 7275 &0.50  &$42  \pm10$  &u          \nl
          & \nev\  & 6690 &0.05  &$4   \pm1 $  &u          \nl
          & \mgii\ & 5465 &1.09\rlap{w}  &$69  \pm10$ &10.7       \nl
          & \ciii\ & 3722 &1.61\rlap{e}  &$74  \pm18$  &12.9 \nl \nl
1303$-$250& \oiii\ & 8620 &0.59  &$59  \pm5 $  &u          \nl
          & \hb\   & 8471 &1.0   &$86  \pm8 $  &8.1       \nl
          & \hg\   & 7587 &0.38  &$33  \pm4 $  &\nodata   \nl
          & \oii\  & 6479 &0.08  &$4.0 \pm0.5$ &u        \nl
          & \nev\  & 5947 &0.09  &$4.0 \pm0.5$ &u          \nl
          & \mgii\ & 4877 &4.17  &$121 \pm12 $ &9.7        \nl \nl
1311$-$270& \mgii\ & 8933\rlap{:}&0.93  &$23 \pm4$   &6.2        \nl
          & \ciii\ & 6094 &1.0   &$18  \pm2$   &6.4       \nl
          & \civ\  & 4951 &2.26  &$34  \pm2$   &11.2       \nl
          & \soiv\ & 4463 &0.4   &$6   \pm1$   &\nodata      \nl
          & \lya\  & 3881 &6.26  &$79  \pm4$   &12.8       \nl \nl
1327$-$214& \ha\   & 10006&3.4   &$434 \pm38$  &4.8      \nl
          & \oiii\ & 7631 &0.35  &$38  \pm3 $  &u          \nl
          & \hb\   & 7412 &1.0   &$103 \pm9 $  &6.5       \nl
          & \hg\   & 6629 &0.27  &$19  \pm2 $  &\nodata         \nl
          & \oii\  & 5680 &0.02  &$1.8 \pm0.3$ &u          \nl
          & \nev\  & 5249 &0.015 &$0.5 \pm0.1$ &u          \nl
          & \mgii\ & 4301 &1.57  &$23  \pm2 $  &12.2       \nl \nl
1349$-$265& \oiii\ & 9622 &\nodata&\nodata&\nodata        \nl
          & \hb\   & 9378 &1.0\rlap{s}   &$160 \pm20$  &8.2    \nl
          & \hg\   & 8383 &0.26  &$39  \pm7 $  &\nodata          \nl
          & \oii\  & 7175 &0.012 &$1.8 \pm0.6$ &u          \nl
          & \nev\  & 6603 &0.05  &$0.7 \pm0.1$ &u          \nl
          & \mgii\ & 5385 &0.26\rlap{a}  &$47  \pm7 $  &10.4     \nl \nl
1351$-$211& \oii\  & 8436 &0.75  &$23  \pm4 $  &u          \nl
          & \nev\  & 7779 &0.12  &$6   \pm0.7$ &u          \nl
          & \mgii\ & 6349 &2.6   &$44  \pm4 $  &11.8       \nl
          & \ciii\ & 4325 &1.0   &$17  \pm5 $  &16.7       \nl \nl
1355$-$215& \oii\  & 9721 &0.65  &$20  \pm2 $  &u         \nl
          & \mgii\ & 7307 &1.48  &$40  \pm5 $  &5.5        \nl
          & \ciii\ & 4970 &1.0   &$27  \pm8 $  &10.2       \nl
          & \civ\  & 4023 &5.33  &$71  \pm8 $  &11.8      \nl \nl
1355$-$236& \oiii\ & 9091 &1.93  &$104 \pm8 $  &u          \nl
          & \hb\   & 8916 &1.0   &$60  \pm12$  &9.7       \nl
          & \hg\   & 7956 &0.57  &$29  \pm10$  &\nodata          \nl
          & \oii\  & 6836 &0.23  &$8   \pm1 $  &u          \nl
          & \nev\  & 6277 &0.08  &$3.3 \pm0.5$ &u          \nl
          & \mgii\ & 5153 &1.36  &$38  \pm7 $  &8.9        \nl \nl
1359$-$281& \oiii\ & 8931 &3.0   &$273 \pm24$  &u          \nl
          & \hb\   & 8767 &1.0   &$91  \pm8 $  &3.7        \nl
          & \hg\   & 7824 &0.34  &$29  \pm3 $  &\nodata       \nl
          & \oii\  & 6715 &0.91  &$74  \pm7 $  &u          \nl
          & \nev\  & 6166 &0.21  &$15  \pm2 $  &u          \nl
          & \mgii\ & 5050 &1.90  &$96  \pm11$  &8.6        \nl \nl
2021$-$208& \mgii\ & 6437 &1.16  &$61  \pm7 $  &13.2      \nl
          & \ciii\ & 4390 &1.0   &$47  \pm5 $  &11.8       \nl \nl
2024$-$217& \ha\   & 9590 &7.47  &$508 \pm61$  &9.0      \nl
          & \oiii\ & 7323 &3.12  &$196 \pm18$  &u          \nl
          & \hb\   & 7090\rlap{:}&1.0 &$63 \pm13$  &8.7       \nl
          & \oii\  & 5459 &0.43  &$41  \pm6$   &u          \nl
          & \nev\  & 5003 &0.06  &$5.5 \pm0.6$ &u          \nl
          & \mgii\ & 4082 &2.12  &$200 \pm44$  &13.0       \nl
          & \ciii\ & [2]~ &\nodata&(12.5)&\nodata       \nl \nl
2025$-$206& \mgii\ & 6720  1.6   &$83  \pm35$  &5.7        \nl
          & \ciii\ & 4582 &1.0   &$29  \pm4 $  &8.0       \nl
          & \civ\  & 3720 &6.5   &$158 \pm27$  &16.4      \nl \nl
2030$-$230& \ha\   & 7430\rlap{:}&9.9   &$593 \pm140$ &7.3    \nl
          &\oiii\  & 5670\rlap{:}&2.45  &$147 \pm12 $ &u          \nl
          & \hb\   & 5500\rlap{:}&1.0   &$60  \pm12 $ &4.9        \nl
          & \oii\  & 4210\rlap{:}&0.93  &$56  \pm7  $ &u          \nl \nl
2035$-$203& \ha    & 9941 &4.8\rlap{s}   &$521 \pm44$  &7.5     \nl   
          & \oiii  & 7592 &0.4   &$27  \pm3 $  &u      \nl
          & \hb    & 7372 &1.0\rlap{s}   &$66  \pm7 $  &10.0   \nl
          & \hg    & 6574 &0.14  &$9   \pm1 $  &\nodata       \nl
          & \oii   & 5651 &0.04  &$2.4 \pm0.3$ &u       \nl
          & \mgii  & 4234 &0.7   &$29  \pm3 $  &13.6   \nl \nl
2040$-$236& \oiii  & 8530 &0.4   &$34  \pm3 $  &u      \nl 
          & \hb    & 8279 &1.0   &$84  \pm8 $  &9.3   \nl
          & \hg    & 7383 &0.15  &$20  \pm2 $  &\nodata       \nl
          & \oii   & 6356 &0.01  &$0.4 \pm0.2$ &u       \nl
          & \mgii  & 4763 &1.6   &$39  \pm4 $  &9.4    \nl \nl
2111$-$259& \mgii\ & 4486 &\nodata &$9   \pm2 $  &4.6       \nl \nl
2122$-$238& \mgii\ & 7716 &0.79\rlap{a}  &$8.6 \pm1.0$ &2.9     \nl
          & \ciii\ & 5343 &1.0\rlap{a}   &$7.4 \pm1.0$ &4.7    \nl
          & \civ\  & 4324 &4.7\rlap{a}   &$26  \pm2 $  &6.8    \nl
          & \soiv\ & 3894 &2.8   &$10  \pm1 $  &\nodata         \nl
          & \lya\  & 3377 &8.8\rlap{a}   &$38  \pm2 $  &9.6   \nl \nl
2136$-$251& \oiii\ & 9431 &\nodata&\nodata&\nodata      \nl
          & \hb\   & 8860 &1.0   &$146 \pm54$  &5.2        \nl
          & \hg\   & 8425 &0.36  &$34  \pm7 $  &\nodata      \nl
          & \oii\  & 7228 &0.06  &$7.4 \pm0.6$ &u          \nl
          & \nev\  & 6868 &0.02  &$1.4 \pm0.3$ &u          \nl
          & \mgii\ & 5448 &0.4   &$42  \pm15$  &6.9        \nl \nl
2149$-$200& \ha\   & 9368 &7.4   &$547 \pm72$  &5.3       \nl
          & \oiii\ & 7134 &3.4   &$232 \pm7 $  &u          \nl
          & \hb\   & 6932 &1.0   &$67  \pm6 $  &3.6        \nl
          & \hg\   & 6189 &0.5   &$32  \pm5 $  &\nodata          \nl
          & \oii\  & 5313 &0.6   &$34  \pm3 $  &u          \nl
          & \nev\  & 4883 &0.17  &$6.0 \pm0.6$ &u          \nl
          & \mgii\ & 3994 &1.3   &$62  \pm20$  &13.3      \nl \nl
2156$-$245& \oiii\ & 9368 &\nodata&\nodata&\nodata       \nl
          & \hb\   & 9050 &1.0\rlap{s}   &$30  \pm15$  &4.4       \nl
          & \hg\   & 8096 &0.4   &$12  \pm5 $  &\nodata          \nl
          & \oii\  & 6952 &0.5\rlap{n}   &$12  \pm2 $  &u      \nl
          & \nev\  & 6370 &0.2   &$4   \pm2 $  &u          \nl
          & \mgii\ & 5215 &0.3\rlap{a}   &$15  \pm3 $  &9.0   \nl \nl
2158$-$206& \mgii\ & 9117 &0.30  &$30  \pm2 $  &3.1        \nl
          & \ciii\ & 6188 &1.0   &$60  \pm2 $  &8.2       \nl
          & \civ\  & 5106 &0.72  &$35  \pm4 $  &6.4       \nl
          & \soiv\ & 4573 &0.85  &$16  \pm2 $  &\nodata      \nl
          & \lya\  & 3998 &2.9   &$94  \pm9 $  &11.3       \nl \nl
2210$-$257& \mgii\ & 7938 &0.5   &$4.4 \pm1.0$ &3.5        \nl
          & \ciii\ & 5406 &1.0   &$7   \pm1 $  &5.0       \nl
          & \civ\  & 4400 &5.7   &$33  \pm3 $  &7.2       \nl
          & \lya\  & [2]~ &\nodata&(120)&\nodata    \nl \nl
2211$-$251& \mgii\ & 9800\rlap{:}&\nodata&\nodata&\nodata        \nl
          & \ciii\ & 6675 &1.0\rlap{n}   &$6.0 \pm0.5$ &2.7\rlap{:}       \nl
          & \civ\  & 5340 &4.6\rlap{n}   &$27  \pm5  $ &1.4\rlap{:}       \nl
          & \lya\  & 4312 &14\rlap{n}    &$65  \pm13 $ &1.7\rlap{:}    \nl \nl
2213$-$283& \oiii\ & 9749 &0.78  &$104 \pm6  $ &u          \nl
          & \hb\   & 9469 &1.0   &$133 \pm7  $ &9.7       \nl
          & \hg\   & 8447 &0.19  &$23  \pm2  $ &\nodata          \nl
          & \oii\  & 7263 &0.05  &$3.8 \pm0.2$ &u          \nl
          & \nev\  & 6668 &0.04  &$4.2 \pm0.2$ &u          \nl
          & \mgii\ & 5437 &1.6   &$83  \pm4  $ &12.6       \nl
          & \ciii\ & 3730 &0.55\rlap{e}  &$20  \pm3  $ &\nodata   \nl \nl
2232$-$272& \oii\  & 9272 &0.11  &$26  \pm7  $ &u         \nl
          & \mgii\ & 6979 &1.4\rlap{a}   &$132 \pm6  $ &12.0       \nl
          & \ciii\ & 4819 &1.0   &$58  \pm9  $ &12.4       \nl
          & \civ\  & 3884 &5.1\rlap{a}   &$225 \pm26 $ &17.6       \nl \nl
2255$-$282& \oiii  & 9632 &0.6   &$36  \pm6  $ &u          \nl 
          & \hb\   & 9359 &1.0   &$59  \pm6  $ &3.6        \nl
          & \hg\   & 8376 &0.9   &$43  \pm4  $ &\nodata        \nl
          & \oii\  & 7176 &0.01  &$0.4 \pm0.1$ &u          \nl
          & \mgii\ & 5374 &1.8\rlap{j}   &$44  \pm8  $ &6.5        \nl
          & \ciii\ & 3670 &1.3\rlap{e}   &$21  \pm4  $ &6.1  \nl \nl
2256$-$217& \mgii\ & 7757 &1.7\rlap{a}   &$34  \pm5  $ &10.4       \nl
          & \ciii\ & 5363 &1.0   &$16  \pm2 $  &9.1       \nl
          & \civ\  & 4312 &2.2\rlap{n}   &$30  \pm7 $  &13.4    \nl
          & \lya\  & 3347 &4.7\rlap{n}   &$53  \pm4 $  &11.6   \nl \nl
2257$-$270& \oii\  & 9220 &0.3   &$15  \pm5 $  &u         \nl
          & \mgii\ & 6952 &1.0   &$36  \pm6 $  &4.0        \nl
          & \ciii\ & 4738 &1.0   &$24  \pm3 $  &4.6       \nl
          & \civ\  & 3849 &5.8   &$128 \pm10$  &6.0       \nl \nl
2338$-$233& \oiii\ & 8605 &1.0   &$69  \pm3 $  &u          \nl
          & \hb\   & 8393 &1.0   &$69  \pm6 $  &10.8       \nl
          & \hg\   & 7482 &0.26  &$15  \pm2 $  &\nodata          \nl
          & \oii\  & 6402 &0.20  &$9   \pm1 $  &u          \nl
          & \nev\  & 5872 &0.09  &$4.0 \pm0.5$ &u          \nl
          & \mgii\ & 4802 &1.8   &$45  \pm3 $  &15.6       \nl \nl
2338$-$290& \ha\   & 9516 &5.5\rlap{s}   &$270 \pm21$  &7.2       \nl
          & \oiii\ & 7249 &0.9   &$35  \pm2 $  &u          \nl
          & \hb\   & 7037 &1.0   &$35  \pm2 $  &9.7      \nl
          & \hg\   & 6317 &0.56  &$18  \pm1 $  &\nodata          \nl
          & \oii\  & 5396 &0.19  &$4.6 \pm0.3$ &u          \nl
          & \nev\  & 4961 &0.15  &$5.0 \pm0.5$ &u          \nl
          & \mgii\ & 4071 &0.31  &$58  \pm4 $  &11.6       \nl \nl
2348$-$252& \oii\  & 8893\rlap{:}&0.006 &$0.4 \pm0.1$ &u          \nl
          & \nev\  & 8176 &0.006 &$0.4 \pm0.1$ &u          \nl
          & \mgii\ & 6676 &0.41  &$19  \pm4  $ &10.6       \nl
          & \ciii\ & 4558 &1.0   &$31  \pm3  $ &13.7      \nl
          & \civ\  & 3723 &1.08  &$24  \pm5  $ &15.9       
\enddata
\end{deluxetable}


\label{ewtab}



\clearpage

\begin{figure*}
\centerline{\psfig{file=f1.eps,bbllx=86pt,bblly=88pt,bburx=508pt,bbury=
370pt,clip=,height=5cm} }
\end{figure*}

\newpage


\newpage

\begin{figure*}
\centerline{\psfig{file=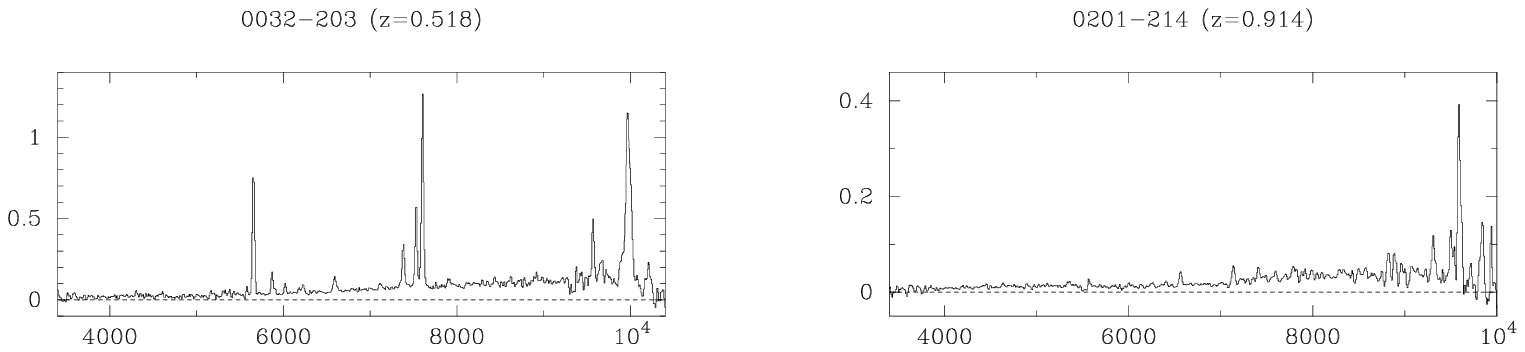,bbllx=86pt,bblly=86pt,bburx=510pt,bbury=
365pt,clip=,height=5.2cm} }
\end{figure*}

\newpage

\begin{figure*}
\centerline{\psfig{file=f4.eps,bbllx=86pt,bblly=18pt,bburx=435pt,bbury=
370pt,clip=,height=7.4cm} }
\end{figure*}

\begin{figure*}
\centerline{\psfig{file=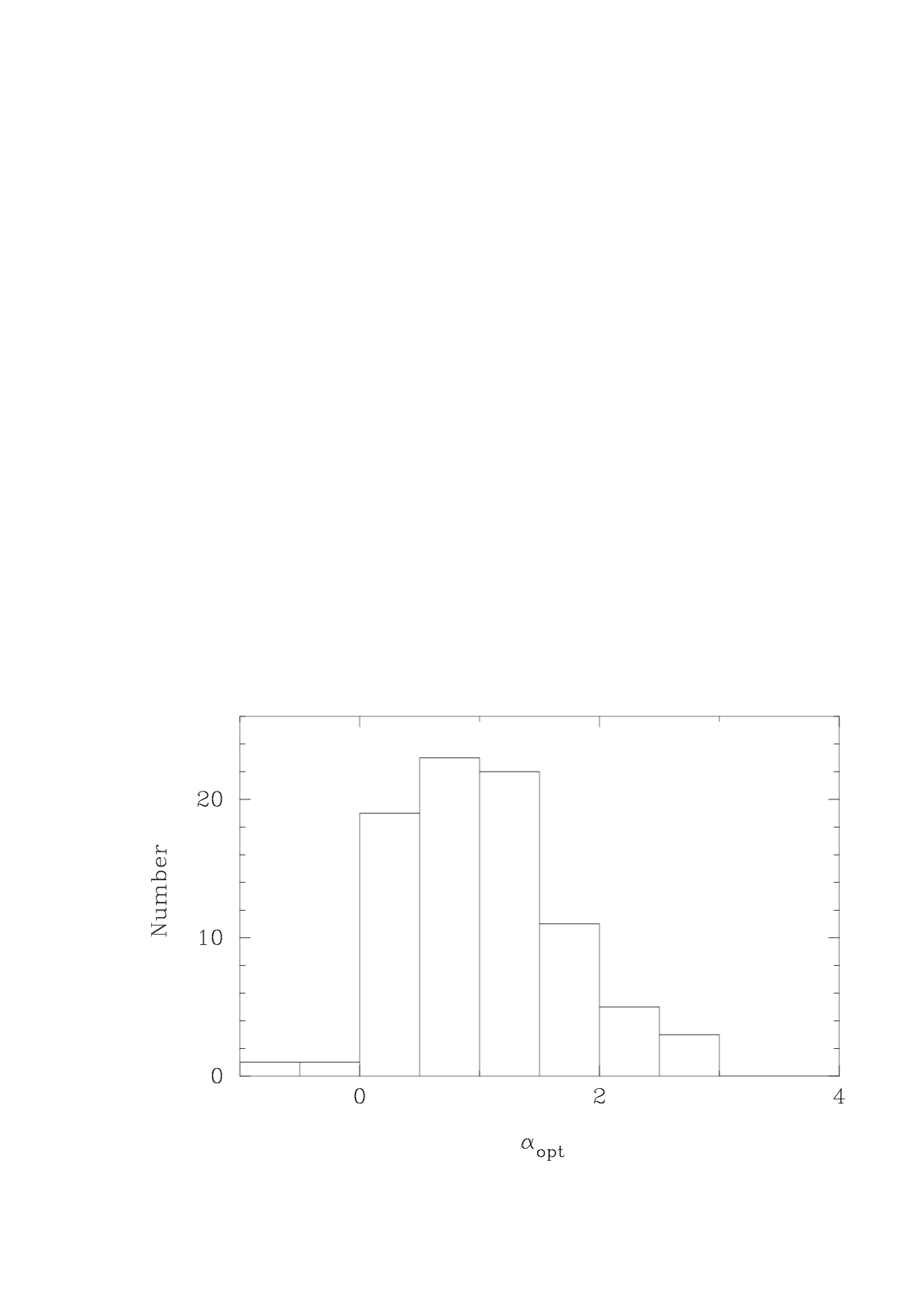} }
\end{figure*}

\begin{figure*}
\centerline{\psfig{file=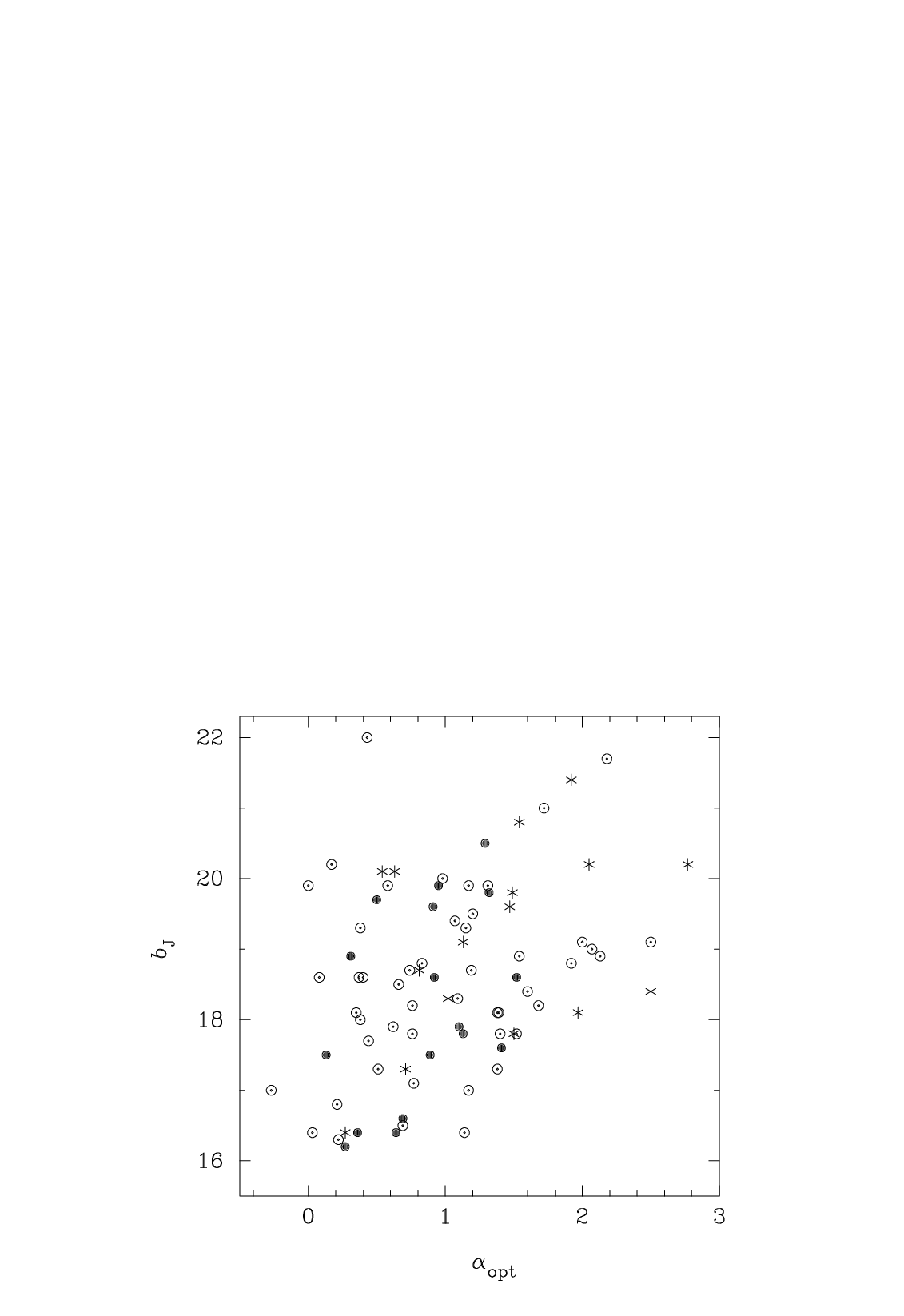}}
\end{figure*}

\begin{figure*}
\centerline{\psfig{file=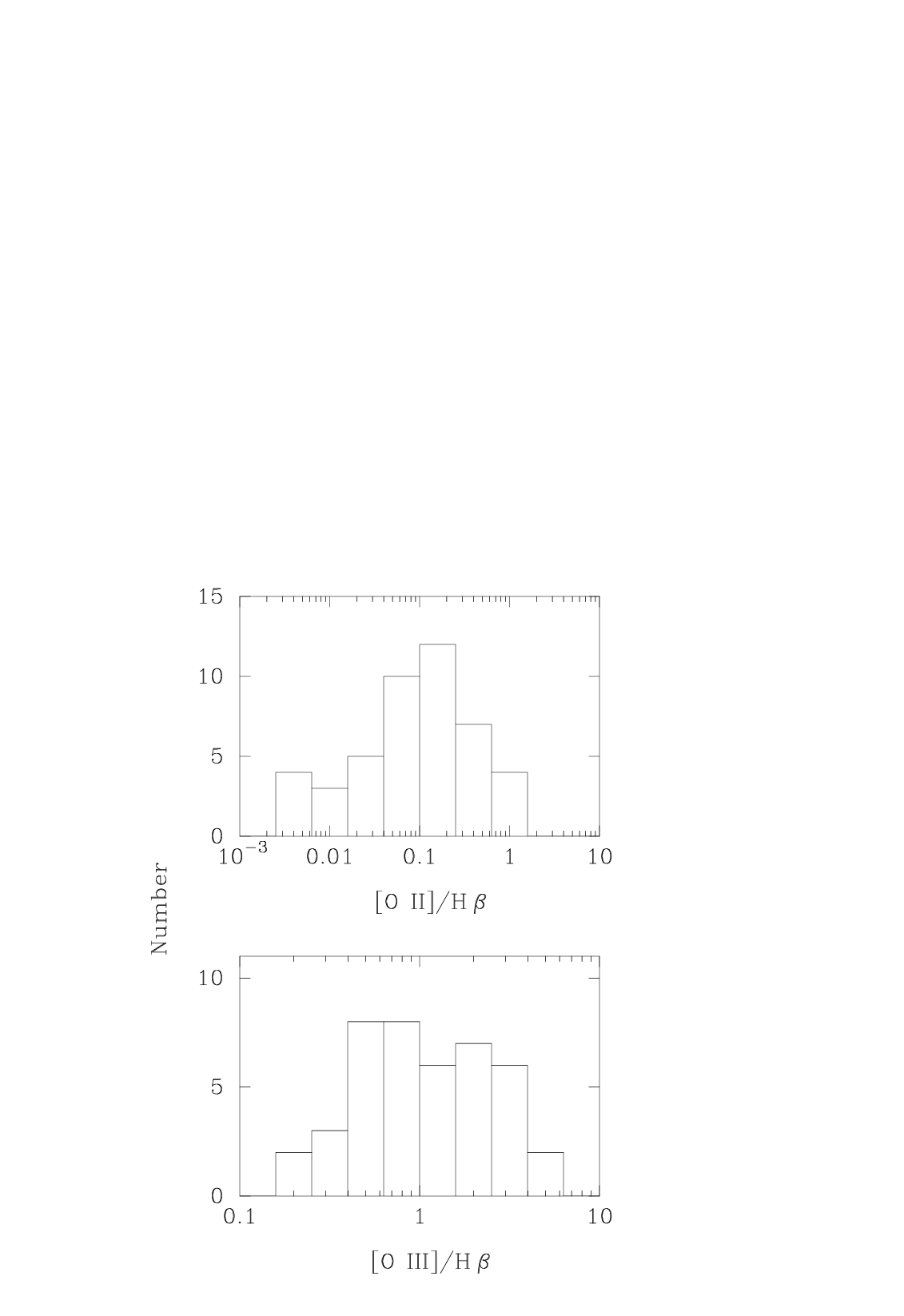,bbllx=90pt,bblly=20pt,bburx=380pt,bbury=
450pt,clip=,height=7.5cm} }
\end{figure*}

\begin{figure*}
\centerline{\psfig{file=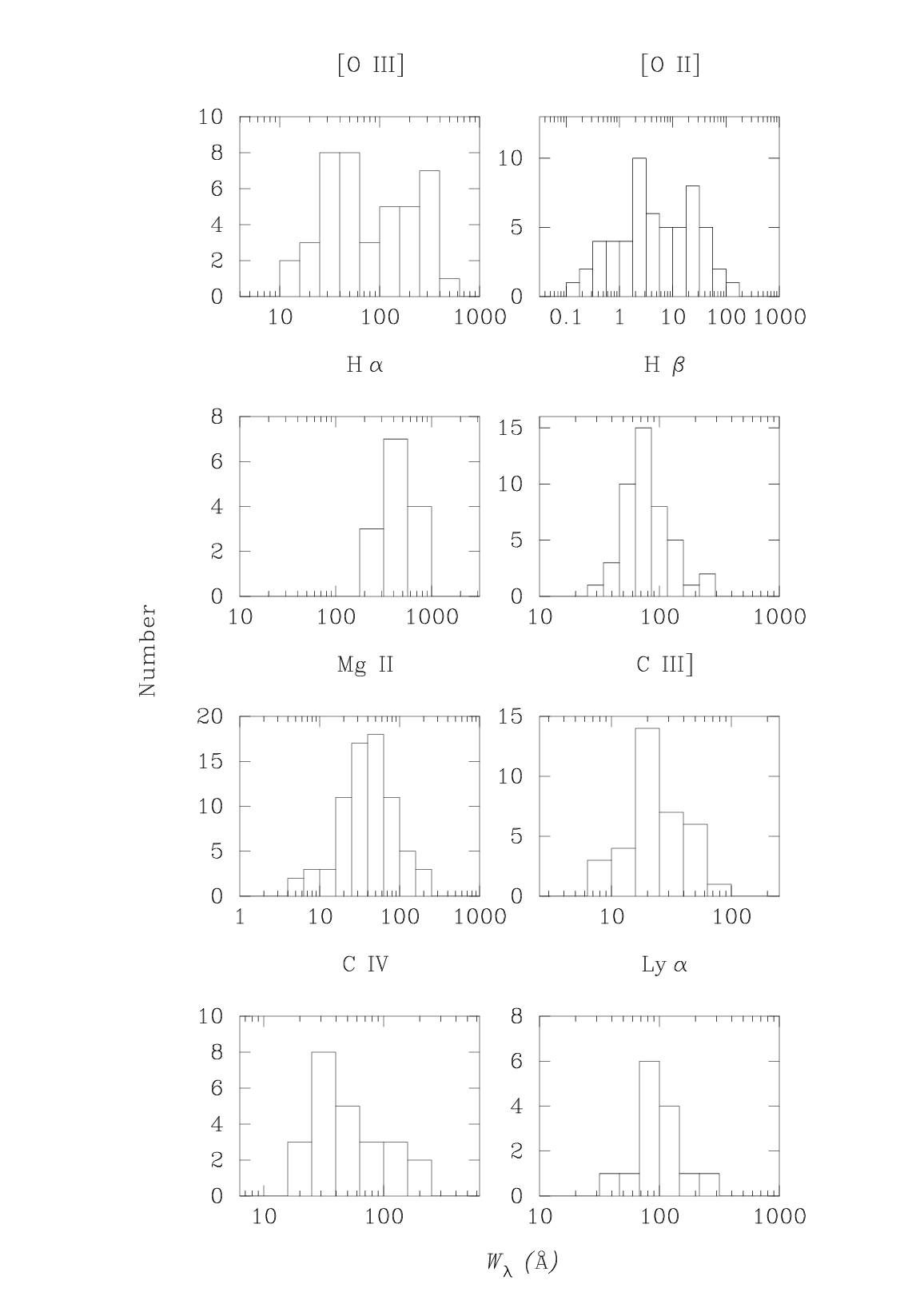,bbllx=84pt,bblly=20pt,bburx=485pt,bbury=
765pt,clip=,height=15cm} }
\end{figure*}

\begin{figure*}
\centerline{\psfig{file=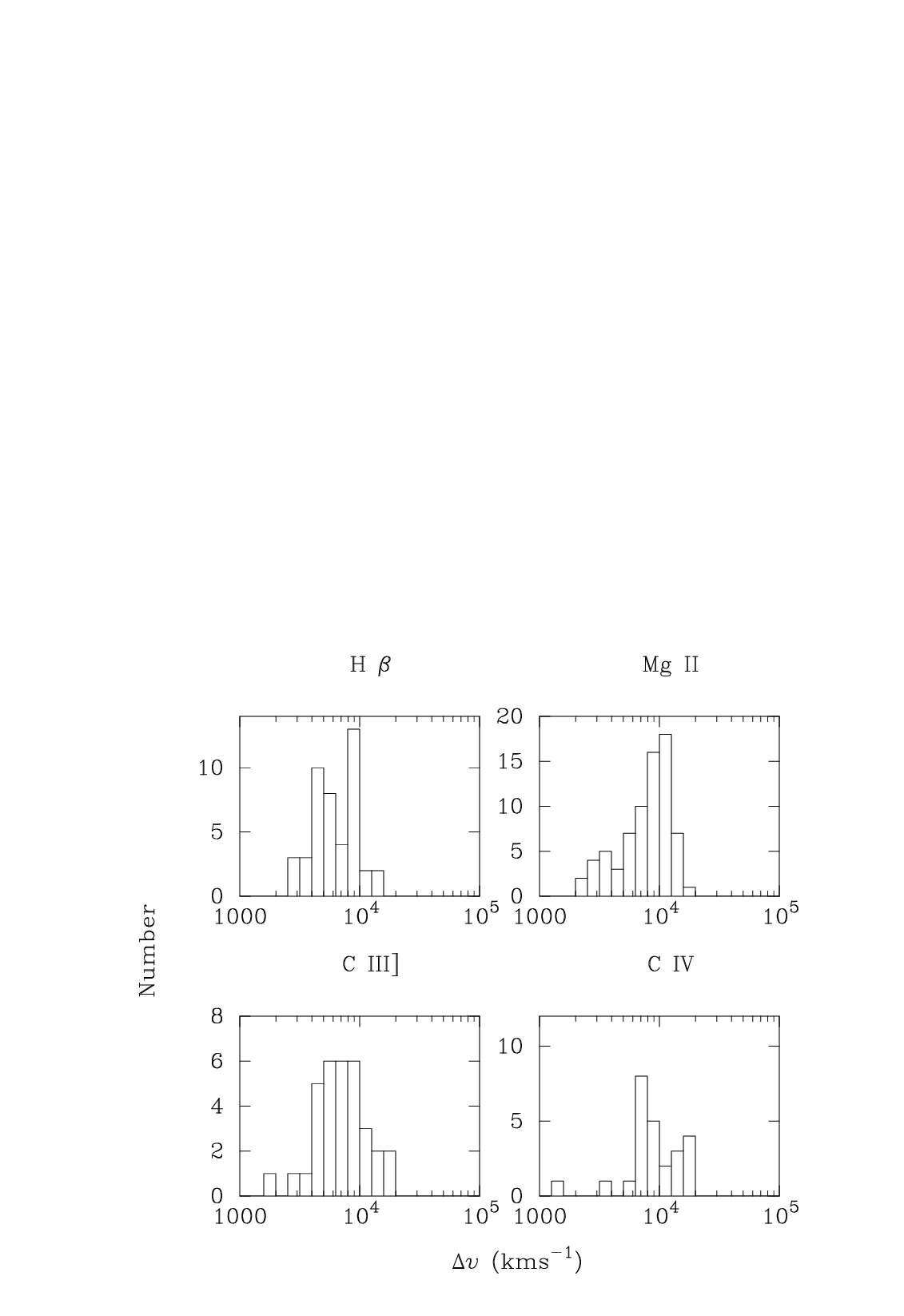,bbllx=84pt,bblly=20pt,bburx=480pt,bbury=
400pt,clip=,height=8cm} }
\end{figure*}

\end{document}